\documentclass[preprint,superscriptaddress,longbibliography]{revtex4-1}
\usepackage{multirow}
\usepackage{color}
\usepackage{bm}
\usepackage{ulem}
\usepackage{graphicx}
\usepackage{mathrsfs}
\usepackage{epsfig,psfrag}
\usepackage{amsmath}
\usepackage{amssymb}
\usepackage{amsbsy}
\usepackage{amsthm}
\usepackage{amsfonts}
\usepackage{wasysym}
\usepackage{bbm}
\usepackage{tabularx}
\usepackage{euscript}
\usepackage{enumerate}
\usepackage{amsfonts}
\usepackage{exscale}
\usepackage{bbold}
\usepackage{float}
\usepackage{slashed}
\usepackage{subfigure}
\usepackage{comment}
\usepackage[]{hyperref}

\def\bea{\begin{eqnarray}}
\def\eea{\end{eqnarray}} 
\def\nn{\nonumber}
\def\ba{\begin{array}} 
	\def\ea{\end{array}}
\def\nn{\nonumber}

\def\sgn{\text{sgn}}

\def\Area{\text{Area}}

\def\kc#1{\left(#1\right)}
\def\kd#1{\left[#1\right]}
\def\ke#1{\left\{#1\right\}}

\def\sgn{{\rm sgn}}

\def\be{\begin{equation}}       \def\ee{\end{equation}}
\def\bea{\begin{eqnarray}}      \def\eea{\end{eqnarray}}
\def\ba{\begin{array}}
	\def\ea{\end{array}}
\def\bnum{\begin{enumerate} }
	\def\enum{\end{enumerate}}

\def\nn{\nonumber}

\def\=>{\Rightarrow}
\def\>{\rightarrow}

\def\eye2{Fathbb{I}}

\DeclareMathOperator\sech{sech}

\begin{document}
	
\title{Reflected Entropy in Double Holography}

\author{Yi Ling}
\email{lingy@ihep.ac.cn}
\affiliation{
Institute of High Energy Physics, Chinese Academy of Sciences, Beijing 100049, China}
\affiliation{School of Physics, University of Chinese Academy of Sciences, Beijing 100049, China}

\author{Peng Liu}
\email{phylp@email.jnu.edu.cn}
\affiliation{Department of Physics and Siyuan Laboratory, Jinan University, Guangzhou 510632, China}

\author{Yuxuan Liu}
\email{liuyuxuan@ucas.ac.cn}
\affiliation{Kavli Institute for Theoretical Sciences (KITS), University of Chinese Academy of Sciences, Beijing 100190, China}
\affiliation{
Institute of High Energy Physics, Chinese Academy of Sciences, Beijing 100049, China}
\affiliation{School of Physics, University of Chinese Academy of Sciences, Beijing 100049, China}

\author{Chao Niu}
\email{niuchaophy@gmail.com}
\affiliation{Department of Physics and Siyuan Laboratory, Jinan University, Guangzhou 510632, China}

\author{Zhuo-Yu Xian}
\email{zhuo-yu.xian@physik.uni-wuerzburg.de}
\affiliation{
Institute for Theoretical Physics and Astrophysics and W\"urzburg-Dresden Cluster of Excellence ct.qmat, Julius-Maximilians-Universit\"at W\"urzburg, 97074 W\"urzburg, Germany}
\affiliation{Institute of Theoretical Physics, Chinese Academy of Science, Beijing 100190, China}

\author{Cheng-Yong Zhang}
\email{zhangcy@email.jnu.edu.cn}
\affiliation{Department of Physics and Siyuan Laboratory, Jinan University, Guangzhou 510632, China}

\begin{abstract}
	Recently, the reflected entropy is proposed in holographic approach to describe the entanglement of a bipartite quantum system in a mixed state, which is identified as the area of the reflected minimal surface inside the entanglement wedge. In this paper, we study the reflected entropy in the doubly holographic setup, which contains the degrees of freedom of quantum matter in the bulk. In this context, we propose a notion of quantum entanglement wedge cross-section, which may describe the reflected entropy with higher-order quantum corrections. We numerically compute the reflected entropy in pure AdS background and black hole background in four dimensions, respectively. In general, the reflected entropy contains the contribution from the geometry on the brane and the contribution from the CFT. We compute their proportion for different Newton constants and find that their behaviors are in agreement with the results based on the semi-classical gravity and the correlation of CFT coupled to the bath CFT.
\end{abstract}
\maketitle

\tableofcontents

\section{Introduction}
The holographic entanglement entropy (HEE) has provided a geometric description for the entanglement of quantum matter and thus opened a new window for understanding the fundamental problems in quantum information theory. Originally, it is identified with the area of the minimum surface ending on the boundary. When quantum fields in the bulk are taken into account, their contribution to the entanglement can be evaluated by considering the minimal area of the quantum extremal surface (QES). Specifically, given a $d$ dimensional asymptotically AdS spacetime and consider a region $A$ on the boundary, the von Neumann entropy of this region can be computed by \cite{Ryu:2006bv,Faulkner:2013ana,Engelhardt:2014gca,Lewkowycz:2013nqa}
\begin{align}\label{QES}
S(A)=\min_{X_A}\mathop{ext}_{X_A} \kd{\frac{\Area(X_A)}{4G^{(d)}}+S(\Sigma_A)},
\end{align}
where $G^{(d)}$ is the Newton constant of gravity in $d$-dimensional spacetime. $\Area(X_A)$ denotes the area of QES $X_A$, which stretches into the bulk with $A$ as the boundary. $\Sigma_A$ is the spatial region enclosed by $X_A\cup A$, and throughout this paper we will call $\Sigma_A$ the entanglement wedge of $A$. Thus $S(\Sigma_A)$ denotes the entropy of the quantum field within $\Sigma_A$. Finally, the entropy is identified with the minimal area of all possible QESs. Usually, the entanglement entropy of quantum fields is difficult to compute, however, if they are described by conformal field theory (CFT) with large central charge, then they would enjoy the holographic duality such that we may provide a geometric description for their entanglement entropy by holography as well. The strategy is further embedding the considered $d$-dimensional spacetime into a $d+1$-dimensional spacetime and treating it as a dynamical brane living in the bulk or on the boundary. This setup is also dubbed as double holography. By virtue of this setup, both terms in equation (\ref{QES}) have a geometrical interpretation and the formula becomes \cite{Almheiri:2019hni,Almheiri:2019yqk,Chen:2020uac}
\begin{align}\label{EEDHolo}
S(A)=\min_{X_A}\mathop{ext}_{X_A} \kd{\frac{\Area(X_A)}{4 G^{(d)}_b}+\frac{\Area(X_{\Sigma_A})}{4G^{(d+1)}}},
\end{align} 
where $G^{(d+1)}$ is the $d+1$-dimensional Newton constant and $G^{(d)}_b$ is the intrinsic Newton constant on the brane. Now, thanks to the notion of HEE, $X_{\Sigma_A}$ is identified as the minimal surface associated with the entanglement wedge $\Sigma_{A}$ in the $(d+1)$-dimensional bulk, which may simply be called the Ryu-Takayanagi (RT) surface of $\Sigma_A$. 

\begin{figure}
\subfigure[]{
    \includegraphics[height=0.2\linewidth]{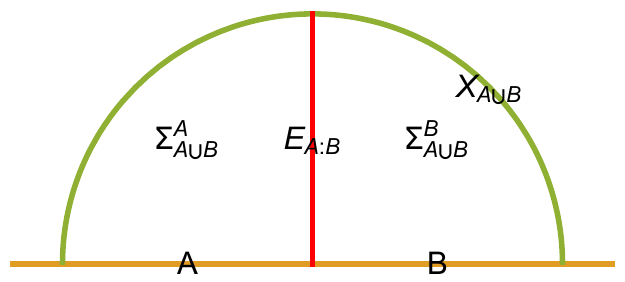}
	\label{fig:EWCS2D}}
\subfigure[]{
    \includegraphics[height=0.26\linewidth]{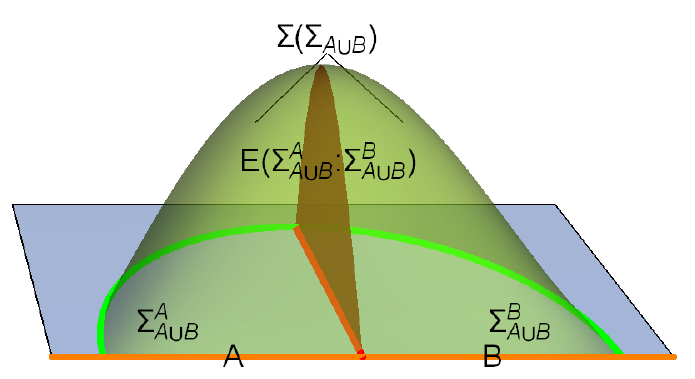}
    \label{fig:EWCS3D}}
	\caption{(a) The cartoon of an EWCS $E_{A:B}$ in the holography of AdS$_3$, where a bipartite system with subregion A and B is set on the boundary. (b) The cartoon of the EWCS in double holography, where AdS$_3$ as a brane (plotted in blue) is embedded into a 4-dimensional spacetime. The reflected entropy is contributed by two terms, namely the area of $E_{A:B}$ and the area of $E\kc{\Sigma_{A\cup B}^A:\Sigma_{A\cup B}^B}$.}
\end{figure}

When a bipartite quantum system with two subregions $A$ and $B$ is in a mixed state, the above setup can be generalized to consider the entanglement between $A$ and $B$ by purification. It has been conjectured that the holographic entanglement of purification could be evaluated by the area of the minimal cross-section of the entanglement wedge (EWCS), which may be denoted as $E_{A:B}$ \cite{Takayanagi:2017knl}. It is expected that this identification captures both classical and quantum correlations between two disjoint subregions. Meanwhile, a similar concept called holographic reflected entropy, which describes the entanglement involving the canonical purification of mixed states, has also been related to the EWCS \cite{Dutta:2019gen}. 
EWCS, as a good measure of mixed state entanglement, has been widely studied in recent literature \cite{Umemoto:2018jpc,Yang:2018gfq,Kudler-Flam:2018qjo,Kusuki:2019zsp,Dutta:2019gen,Huang:2019zph,Fu:2020oep,Gong:2020pse,Liu:2019qje,Lala:2020lcp,Bao:2018gck,Bueno:2020fle,KumarBasak:2020eia}.
Similar to the holographic dual of entanglement entropy, the holographic dual of reflected entropy with quantum fields in the bulk is proposed as \cite{Dutta:2019gen}
\begin{align}\label{EWCS}
S^R(A:B)=\min_{E_{A:B}} \kd{\frac{\Area(E_{A:B})}{4G^{(d)}}}+S^R(\Sigma_{A\cup B}^A:\Sigma_{A\cup B}^B)|_{E_{A:B}^{\min}}+O(G^{(d)}),
\end{align}
where the first term is proportional to the area of the EWCS $E_{A:B}$ that splits the wedge $\Sigma_{A\cup B}$ into two parts and the second term is the reflected entropy between the quantum fields in the bipartition $\Sigma_{A\cup B}^A:\Sigma_{A\cup B}^B$, as illustrated in Fig.~\ref{fig:EWCS2D}. In this figure, one intuitively notices that   $\Sigma_{A\cup B}=\Sigma_{A\cup B}^A\cup\Sigma_{A\cup B}^B$ and $E_{A:B}=\Sigma_{A\cup B}^A\cap\Sigma_{A\cup B}^B$. 
Next, for convenience,  we call the second term the bulk reflected entropy.

Similar to the arguments on the
holographic entanglement entropy in \cite{Faulkner:2013ana}, the holographic reflected entropy in (\ref{EWCS}) does not contain quantum corrections $o(G^{(d)}{}^0)$. In \cite{Faulkner:2013ana}, a very elegant scheme has been proposed to include the contribution of quantum corrections of HEE. The key point is to extend the notion of extremal surface to quantum extremal surface, which is obtained by finding the minimal contribution of EE from both terms, as shown in (\ref{QES}). Motivated by this point, we propose a generalization of EWCS to its quantum version such that the holographic reflected entropy contains higher-order quantum corrections as well in this paper. Specifically, in the presence of quantum fields in the bulk, we propose that the reflected entropy between $A$ and $B$ on the boundary can be evaluated by holography as
\begin{align}\label{QEWCS}
S^R(A:B)=\min_{E_{A:B}} \kd{\frac{\Area(E_{A:B})}{4G^{(d)}}+S^R(\Sigma_{A\cup B}^A:\Sigma_{A\cup B}^B)}.
\end{align}
In comparison with the equation in (\ref{EWCS}), the key difference is that searching the minimum is taken at the final step such that the minimal cross-section $E_{A:B}^{\min}$ is influenced by the entanglement between the quantum fields in the bulk regions $\Sigma_{A\cup B}^A$ and $\Sigma_{A\cup B}^B$ as well. So we call it quantum entanglement wedge cross-section (QEWCS). Obviously, when the total system $A\cup B$ is in a pure state, the holographic reflected entropy (\ref{QEWCS}) recovers the holographic entanglement entropy in (\ref{QES}) and the QEWCS recovers the QES. However, in general mixed states, we are usually stuck by the difficulty of computing the entanglement of quantum fields, which is the second term in (\ref{QEWCS}). 
To overcome this difficulty, we intend to investigate the reflected entropy with quantum corrections by virtue of the doubly holographic setup. 

The reflected entropy was previously studied in some doubly holographic setups, focusing on the island scenario of reflected entropy \cite{Chandrasekaran:2020qtn,Li:2020ceg}. The EWCS of the reflected entropy in the $(d+1)$-dimensional spacetime may end either on the $(d-1)$-dimensional RT surface in the $(d+1)$-dimensional spacetime or on the $d$-dimensional brane, where the holographic reflected entropy of some regions on the $d$-dimensional boundary may contain the geometric contribution of the island in the dynamical spacetime on the $d$-dimensional brane theory. In contrast to the above consideration, we will utilize the double holography in a quite different way, where both subregions $A$ and $B$ are located on the conformal boundary of $d$-dimensional spacetime.
In the doubly holographic setup, we propose that the reflected entropy with quantum corrections in (\ref{QEWCS}) can be evaluated by the following formula
\begin{align}\label{REDHolo}
S^R(A:B)=\min_{E_{A:B}} \kd{\frac{\Area(E_{A:B})}{4G_b^{(d)}}+\frac{\Area\kd{E\kc{\Sigma_{A\cup B}^A:\Sigma_{A\cup B}^B}}}{4G^{(d+1)}}},
\end{align}
where $E\kc{\Sigma_{A\cup B}^A:\Sigma_{A\cup B}^B}$ is the EWCS that splits the entanglement wedge of $\Sigma_{A\cup B}$, which is denoted as $\Sigma(\Sigma_{A\cup B})$, into two parts in the $(d+1)$-dimensional spacetime.  We illustrate the cartoon of the EWCS in double holography in Fig.~\ref{fig:EWCS3D}. 

Equation (\ref{REDHolo}) is the core formula proposed in the present paper. Next, we will present the details for the doubly holographic setup, and then evaluate the reflected entropy with quantum corrections for some  bipartite systems in pure AdS space and black hole background, respectively.

\section{The doubly-holographic setup}

Consider a $d$-dimensional Planck brane $Q$ living in a $(d+1)$-dimensional asymptotic AdS space $N$, which is called bulk. The brane ends on the conformal boundary $M$ of the asymptotic AdS space and their intersection forms a $(d-1)$-dimensional space $P=M\cap Q$ \cite{Almheiri:2019hni,Chen:2020uac,Almheiri:2019yqk,Takayanagi:2011zk}. As a result, the full boundary of the asymptotic AdS space becomes $\partial N=M\cup Q$. We consider an action of the brane, which contains a tension term and Dvali-Gabadadze-Porrati (DGP) term \cite{Randall:1999vf,Dvali:2000hr}. So the total action of the system is given as 
\begin{align}\label{ActionDHolo}
I=&\frac{1}{16\pi G^{(d+1)}} \bigg [ \int_N d^{d+1}x
\sqrt{-g}\kc{R+\frac{d(d-1)}{L^2}} \nn \\
&+2\int_{M}d^{d}x\sqrt{-h}K 
+2\int_{Q}d^{d}x\sqrt{-h}(K-\alpha)
-2\int_P \sqrt{-\Sigma}\vartheta \bigg ]   \\
&+ \frac{1}{16 \pi G_{DGP}^{(d)}} \kd{ \int_Q d^d x \sqrt{-h}R_h
	+2\int_P \sqrt{-\Sigma} k} \nn,
\end{align}
where $h$ is the induced metric on the boundary, and $\Sigma$ the induced metric on $P$. $K$ is its extrinsic curvature scalar and $R_h$ is the intrinsic curvature scalar of $h$. The $\vartheta$ and $k$ are the intrinsic curvature scalar and extrinsic curvature scalar of $P$. The constant $\alpha$ is proportional to the tension of the brane and for simplicity we just call it tension term. The third line in (\ref{ActionDHolo}) is DGP term where a $d$-dimensional Newton constant $G_{DGP}^{(d)}$ is introduced. 

To determine the metric of the background, we need to solve the equations of motion. For this purpose, we impose Dirichlet boundary condition on the conformal boundary $M$ and Neumann boundary condition on the brane $Q$
\begin{align}
M:&\quad h_{ij}=\frac1{\epsilon^2}\eta_{ij},\\
Q:&\quad  K_{ij}-K h_{ij}+\alpha h_{ij}=\lambda L \left[\frac{1}{2}R_h h_{ij}-(R_{h}){}_{ij}\right], \quad \lambda=G^{(d+1)}/(G_{DGP}^{(d)}L), \label{eq:BCBrane}
\end{align}
where $\epsilon$ is the length cutoff of the theory on the conformal boundary. We will take the semi-classical limit $L^{d-1}/G^{(d+1)}\to\infty$ such that the background is described by the classical solutions to the Einstein equation on $N$. 

The above system can be viewed from the following three perspectives \cite{Almheiri:2019hni,Chen:2020uac}:
\begin{description}
    \item [Bulk perspective] The pure gravity theory in the asymptotic AdS space $N$ with the above boundary conditions on conformal boundary $M$ and the brane $Q$. 
    \item [Brane perspective] The gravity near the brane $Q$ is localized by the $(d+1)$-dimensional negative curvature \cite{Karch:2000ct}. After imposing the Einstein equation in $N$, the theory in the bulk is dual to the theory of induced metric on the brane $Q$ and the CFT living on both $Q$ and $M$ \cite{Gubser:1999vj}. One may think of it as the gravity-plus-CFT theory on $Q$ coupled to the CFT on the flat half space $M$ at the intersection $P$, where the former is the system that we are interested in and the latter may be treated as a bath.
    \item [Boundary perspective] The geometry on the brane is also an asymptotic AdS space. The gravity-plus-CFT theory is dual to the $(d-1)$-dimensional theory without gravity on its boundary, namely, the intersection $P$ \cite{Almheiri:2019hni}. With the language of boundary conformal field theory (BCFT) \cite{Takayanagi:2011zk}, the intersection $P$ is the boundary of the CFT on $M$, and the theory on $P$ forms a conformal defect \cite{Chen:2020uac}.
\end{description}

The gravity-plus-CFT theory in the brane perspective exhibits the following advantages in the study of the reflected entropy (\ref{QEWCS}). 
\begin{itemize}
    \item The CFT has a semi-classical gravity duality characterized by large central charge and entropy. 
    \item The quantity in the square bracket in (\ref{QEWCS}) can be computed by the RT formula in $(d+1)$-dimensional bulk, with the form of that in (\ref{REDHolo}).
    \item The state in the gravity-plus-CFT theory on $Q$ is mixed, caused by its interaction with the bath CFT.
\end{itemize}

Next, we will consider AdS space and black hole as two specific states of the bulk $N$ and compute the reflected entropy of a simple bipartite of region $P$. 

\section{The reflected entropy in AdS space}

\subsection{Background}

The AdS spacetime with a brane is considered as the ground state. We first consider the bulk $N$ metric as AdS$_{d+1}$ spacetime
\begin{align}\label{Slide}
ds^2_N=&L^2\kc{d\rho^2+\cosh^2\rho \cdot \frac{-dt^2+d\zeta^2+d\vec y^2}{\zeta^2}},
\quad  -\infty<\rho<\rho_0.
\end{align}
with the conformal boundary $M$ and the brane $Q$ at 
\begin{align}
M:&\quad \rho=-\infty,\\
Q:&\quad \rho=\rho_0.
\end{align}
From (\ref{Slide}), the induced metric on the brane $Q$ is $AdS_{d}$ spacetime. 

The Neumann boundary condition on the brane gives rise to 
\begin{align}\label{Tension}
\alpha L +\lambda\sech^2\rho_0-2 \tanh\rho_0=0,
\end{align}
which should be satisfied by the above geometry and embedding.

For later convenience, we apply the coordinate transformation
\begin{align}\label{CoordinateRelation}
    z=\zeta\sech\rho,\quad x=-\zeta\tanh\rho
\end{align}
and rewrite the metric in Poincare coordinate system $(z,x,\vec y,t)$ as
\begin{align}\label{Poincare}
ds^2_N=L^2\frac{-dt^2+dz^2+dx^2+d\vec y^2}{z^2}.
\end{align}
We denote the inner angle between the brane $Q$ and the conformal boundary $M$ as $\pi-\theta$ where $0\leq\theta\leq\pi$, then the location of the brane $Q$ can be described by
\begin{align}\label{Intersect}
z+x\tan\theta=0.
\end{align}
It is easy to see that $\theta$ is related to $\rho_0$ by
$\cot\theta=\sinh\rho_0$ or $\csc\theta=\cosh\rho_0$. Thus the boundary condition in (\ref{Tension}) becomes $\alpha L+\lambda \sin^2\theta-2\cos\theta=0$. In general, one can freely choose the values of $\theta\in[0,\pi]$ and $\lambda\in\mathbb R$, but determine $\alpha L$ by the above equation. Basically, we will consider the case with $0<\theta\leq \pi/2$ such that the boundary entropy is always positive \cite{Takayanagi:2011zk}.

\subsection{RT surface}

Throughout this paper, we only consider time-independent states. So we will work on a specific time slice of $\ke{N,M,Q,P}$ and denote them with the same notations for convenience. 

\begin{figure}
\subfigure[~Island phase]{
	\includegraphics[height=0.35\linewidth]{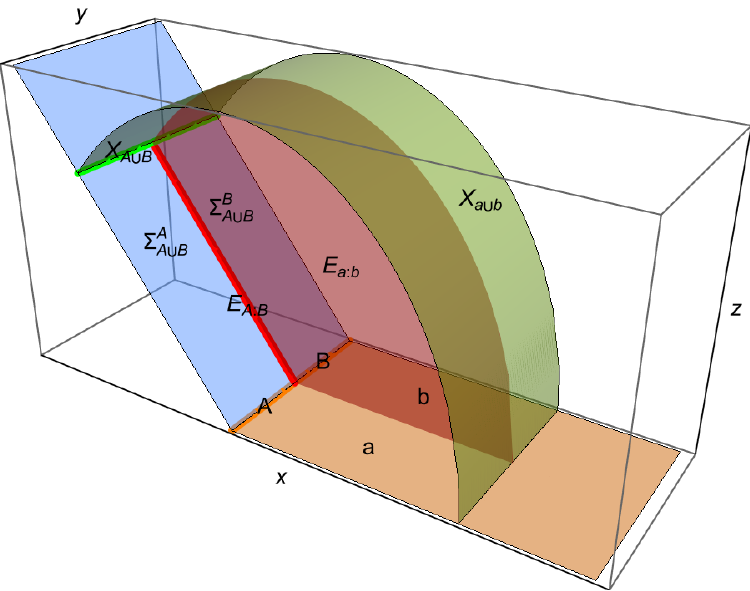}}
\subfigure[~Trivial phase]{
	\includegraphics[height=0.35\linewidth]{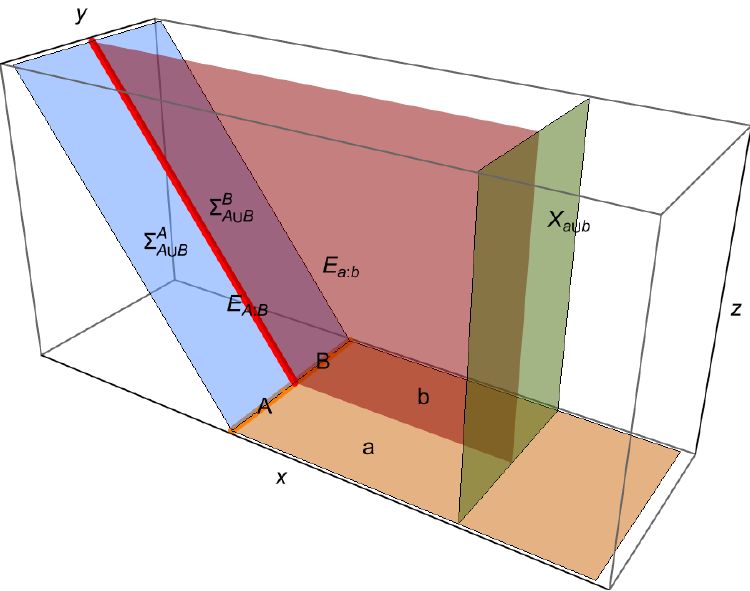}}
	\caption{The two candidates of the RT surface and the EWCS ending on the brane in the double holography of AdS$_4$, where the regions $A$ and $B$ are two half-infinity lines.}
	\label{fig:EWCS3DAdS}
\end{figure}

Rather than considering the bipartition with finite intervals in Fig.~\ref{fig:EWCS3D}, whose entanglement wedge in general is rather complicated for numerical simulations, we will consider the bipartition $A:B$ where $A$ and $B$ are two half-infinite intervals satisfying $P=A\cup B$, as shown in Fig.~\ref{fig:EWCS3DAdS}. In the Poincare patch, we let $n=d-2$, $\vec y=(y,\vec w)$ and $\vec w=(w_1,...,w_{d-3})$. The regions $\ke{P,A,B}$ are defined as
\begin{align}
    P=&\ke{(z,x,y)|z=x=0,y\in\mathbb R}, \nn \\
    A=&\ke{(z,x,y)|z=x=0,y\leq0}, \\
    B=&\ke{(z,x,y)|z=x=0,y\geq0},\nn
\end{align}
which always cover all the space along transverse directions $\vec w$ and their dependence on $\vec w$ has been neglected due to the translational symmetry.

Our goal is to calculate the reflected entropy of $A:B$ by finding its minimal QEWCS. First, we need to figure out the entanglement wedge of $P$. Notice that $P$ is a codimension-$3$ manifold. To apply the RT formula here, we may imagine that $P$ has a finite width along $x$ direction on the boundary $M$ which scales as the UV cutoff $\epsilon$ of the boundary theory. Technically, we will consider a codimension-$2$ region $p=a\cup b\subseteq M$ with bipartition $a:b$ near the brane, which are defined as
\begin{align}
p=&\ke{(z,x,y)|z=0,\ 0\leq x<x_b,\ y\in \mathbb R},\nn\\
a=&\ke{(z,x,y)|z=0,\ 0\leq x<x_b,\ y\leq0},\\
b=&\ke{(z,x,y)|z=0,\ 0\leq x<x_b,\ y\geq 0},\nn
\end{align}
with constant width $x_b$. We can obtain $\ke{P,A,B}$ from $\ke{p,a,b}$ by sending $x_b\to\epsilon$. Thanks to the above limit process, the RT surface of $P$ can be obtained from the RT surface of $p$ by taking the limit. The above setup is illustrated in Fig.~\ref{fig:EWCS3DAdS}. Next, we turn to consider the entanglement in $\ke{p,a,b}$. 

In this subsection, we will focus on the entanglement entropy $S(p)$ associated with the region $p$, but leave the reflected entropy $S^R(a:b)$ for investigation in the next subsection. Now to compute $S(p)$, it is essential to figure out the RT surface $X_p$ and the entanglement wedge $\Sigma_p$ of $p$. 

According to (\ref{EEDHolo}), the entropy $S(p)$ is the minimum in 
\begin{align}\label{EECandidate}
\tilde S(p)=\frac{\lambda L \Area(\tilde X_P)+\Area(\tilde X_p)}{4G^{(d+1)}}
\end{align}
with respect to the surface $\tilde X_p$ anchored on the line $\partial p=\ke{(x,y)|x=x_b,y\in\mathbb R}$ and a line $\tilde X_P=\tilde X_p\cap Q$ on $Q$, where the tildes refer to quantities before minimization.
The minimization can be achieved in two steps. Firstly, given a $\tilde X_P$, we find the minimal surface $X_p$ anchored on $\partial p$. Secondly, we minimize the entropy with respect to $\tilde X_P$ and determine $X_P$. 

In general, there are two candidates of RT surface $X_p$, one of which ends on the brane $Q$ ($X_P\neq\emptyset$) and the other does not ($X_P=\emptyset$), as shown in Fig.~\ref{fig:EWCS3DAdS}. We call the former island phase and the latter trivial phase. The island phase depends on the action on the brane $Q$, while the trivial phase is a surface at $x=x_b$ stretching into the bulk, which is independent from the brane. The entanglement wedge $\Sigma_p$ is the region enclosed by the RT surface and the brane.

We are figuring out the minimal surface $X_p$ at the first step. We work in $\kc{z,x}$ coordinates (\ref{Poincare}) and parameterize $\tilde X_p$ as $(z(x),x)$ or $(z,x(z))$. Then the area of $\tilde X_p$ is proportional to the integral
\begin{align}
\frac{\Area(\tilde X_p)}{L^{n+1}V_{n}z_*^{-n}}
=\int_{\tilde x_0}^{\tilde x_b} d\tilde x \frac{\sqrt{1+\tilde z'(\tilde x)^2}}{\tilde z^{n+1}},\quad  \tilde x=\frac{x}{z_*},\quad   \tilde z=\frac{z}{z_*},
\end{align}
where the undetermined coefficient $z_*$ is the value of $z$ at the turning point $z'(x)=0$. Treating the integral as an action, we find the corresponding equation of motion derived from the Hamiltonian is given by
\begin{align}
1=\frac1{\tilde z^{n+1}\sqrt{1+\tilde z'(\tilde x)^2}}.
\end{align}
$x(z)$ has two solutions $x_\pm(z)$ 
\begin{align}\label{RTAdS}
\tilde x_\pm(\tilde z) =& x_b/z_*-X_n(0) \pm X_n(\tilde z),\\
X_n(\tilde z)=&-i \tilde z \, _2F_1\left(\frac{1}{2},-\frac{1}{2+2n};1-\frac{1}{2+2 n};\tilde z^{-2-2n}\right)
+\frac{i \sqrt{\pi} \Gamma \left(\frac{-1}{2+2n}\right)}{\Gamma\left(-1+\frac{n}{2+2n}\right)}. \nn
\end{align}
The minimal surface $X_p$ parameterized by (\ref{RTAdS}) intersects with $Q$ at $\tilde X_P$. Denote the location of $\tilde X_P$ as $(z_0,x_0)$, which satisfies (\ref{Intersect}).
Then the area of $\tilde X_P$ and $X_p$ are given by
\begin{align}
\Area(X_P)=&L^{n}V_{n} z_0^{-n}
\\
\frac{\Area(\tilde X_p)}{L^{n+1}V_{n}z_*^{-n}}
=&\int_{\tilde x_0}^{\tilde x_b} d\tilde x \frac{\sqrt{1+\tilde z'^2}}{\tilde z^{1+n}}
=\kc{\int_{\tilde\epsilon}^{1}+\sigma\int_{\tilde z_0}^{1}} \frac{d\tilde z}{\tilde z^{n+1}\sqrt{1-\tilde z^{2+2n}}}\\
=&I_n+\frac1{n\tilde\epsilon^{n}}
+\sigma\int_{\tilde z_0}^{1} \frac{d\tilde z}{\tilde z^{n+1}\sqrt{1-\tilde z^{2+2n}}}\\
I_n=&\int_0^1\frac{dz}{z^{n+1}}\kc{\frac1{\sqrt{1-z^{2+2n}}}-1}-\frac1{n}.
\end{align}
where  $V_n=\int d^n\vec y$,  $\sigma=\sgn(x_+(z_*)-x_0)$, $\tilde\epsilon=\epsilon/z_*$ and the $\tilde z(\tilde x)$ is the inverse function of (\ref{RTAdS}).

We are figuring out the location of $X_p$ at the second step. In coordinate system $\kc{\rho,\zeta}$ in (\ref{Slide}), the candidate surface $X_p$ anchored at $\partial p$ and $\tilde X_P$ on both ends can be parameterized as $(\rho,\zeta(\rho))$. As a result, the area of $\tilde X_p$ and $X_P$ are separately given by 
\begin{align}
\frac{\Area(\tilde X_P)}{L^{n}V_{n}}
=&\kc{\frac{\cosh\rho_0}{\zeta(\rho_0)}}^{n},	\\
\frac{\Area(\tilde X_p)}{L^{n+1}V_{n}}
=&\int_{-\infty}^{\rho_0} d\rho \kd{\kc{\frac{\cosh \rho}{\zeta}}^{n}\sqrt{1+\kc{\frac{\zeta'(\rho)\cosh\rho}{\zeta}}^2}}.
\end{align}
So, before the minimization, the dimensionless density of entropy in  (\ref{EECandidate}) is
\begin{align}
\tilde s_p=\frac{4G^{(d+1)}}{L^{n+1}V_{n}}\tilde S(p)
=\lambda \kc{\frac{\cosh\rho_0}{\zeta(\rho_0)}}^{n} + \int_{\rho_\epsilon}^{\rho_0} d\rho \kd{\kc{\frac{\cosh \rho}{\zeta}}^{n}\sqrt{1+\kc{\frac{\zeta'(\rho)\cosh\rho}{\zeta}}^2}}.
\end{align}
By requiring $\delta\tilde s_p/\delta\zeta(\rho)=0$, we obtain the boundary condition of $\zeta(\rho)$ as \cite{Chen:2020uac}
\begin{align}\label{BC}
0=n\lambda-\frac{\zeta '(\rho_0 )\cosh^2\rho_0}{\sqrt{\zeta '(\rho_0 )^2\cosh ^2\rho_0+\zeta (\rho_0)^2}}.
\end{align}
So for the island phase, it is necessary that $n|\lambda|\leq \csc\theta$. But it is not sufficient. We will come back to this point soon.

\begin{figure}
 \subfigure[]{\includegraphics[height=0.25\linewidth]{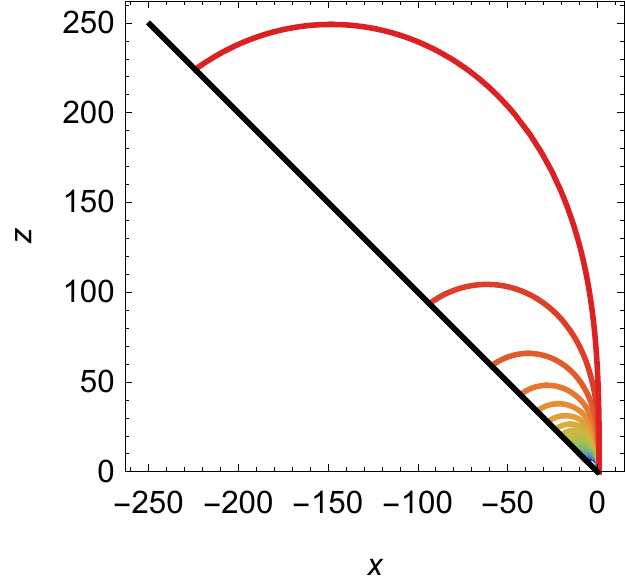}}
 \subfigure[]{\includegraphics[height=0.25\linewidth]{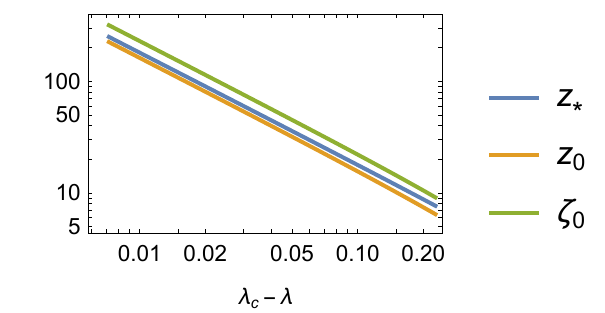}}
 \subfigure[]{\includegraphics[height=0.25\linewidth]{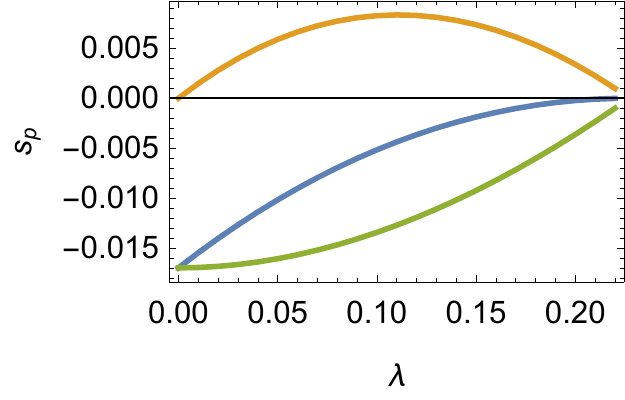}}
 \subfigure[]{\includegraphics[height=0.25\linewidth]{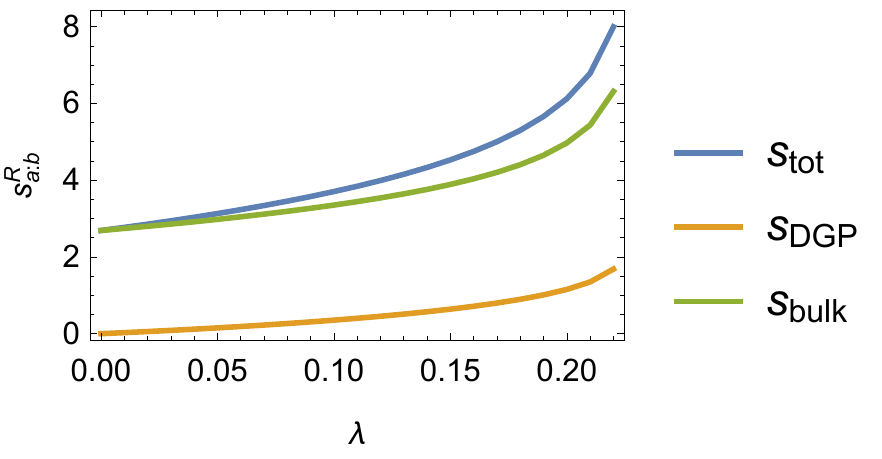}}
	\caption{(a) The RT surface for $\lambda=0, 0.01, 0.02,...,0.22$ (from the bottom to the top) in AdS$_4$. (b) $z_*,\, z_0,\, \zeta_0$ as functions of $\lambda_c-\lambda$. (c)(d) $s_p$ and $s_{a:b}$ as functions of $\lambda$, where $(s_p)_\text{tot}$ and $(s_p)_\text{bulk}$ are renormalized by subtracting $\epsilon^{-n}/n$. The parameters are $d=3,\ \theta=\pi/4,\ x_b=1,\ \epsilon=1$, which determine $\lambda_c=0.227$.}
	\label{fig:RT}
\end{figure}

By utilizing the coordinate relation (\ref{CoordinateRelation}), we can numerically find the value of $z_*$ so that the surface (\ref{RTAdS}) satisfies the boundary condition (\ref{BC}) at the intersection (\ref{Intersect}).
So, the dimensionless entropy density at extremum is given by
\begin{align}\label{EntropyDensityAdS}
s_p=\frac{4G^{(d+1)}}{L^{n+1}V_{n}}S(p)
=\frac{\lambda}{z_0^{n}} + \frac{I_n}{z_*^{n}}+\frac1{n\epsilon^{n}}
+\frac{\sigma}{z_*^{n}}\int_{\tilde z_0}^{1} \frac{d\tilde z}{\tilde z^{n+1}\sqrt{1-\tilde z^{2+2n}}}.
\end{align}
We can numerically check that it is a local minimum. The numerical results for the RT surface $X_p$ and the entanglement entropy density $s_p$ in the island phase are illustrated in Fig.~\ref{fig:RT}. In the trivial phase, the entropy density is simply $s_p=1/(n\epsilon^{n})$, which matches (\ref{EntropyDensityAdS}) at the limit of $z_0,z_*\to\infty$ with finite $\lambda$. 

\begin{figure}
    \centering
    \includegraphics[height=0.25\linewidth]{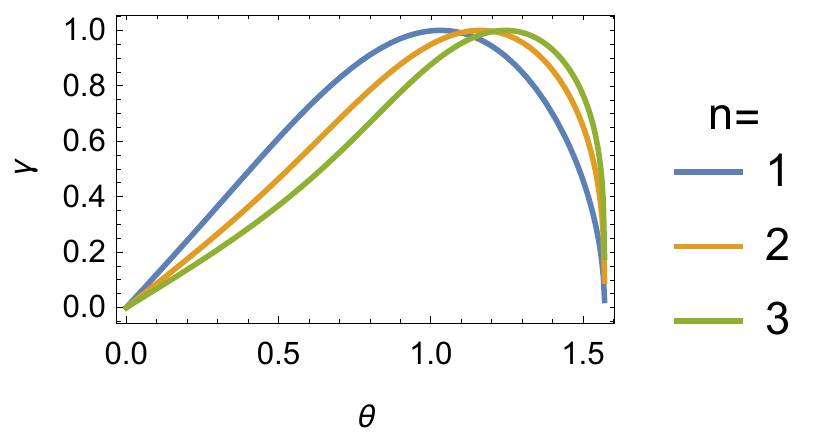}
    \includegraphics[height=0.25\linewidth]{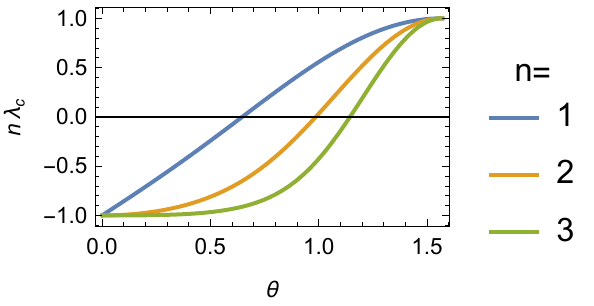}
    \caption{$\gamma$ and $n\lambda_c$ as functions of $\theta$ for $n=1,2,3$.}
    \label{fig:gammalambda}
\end{figure}

Let us compare the two phases for different $\lambda$. Firstly, to avoid the induced gravity on the brane $Q$ becoming unstable \cite{Chen:2020uac}, we require the lower bound $n\lambda>-1$, which is stronger than $n\lambda\geq-\csc\theta$. Secondly, when $n\lambda$ is slightly above $-1$, the island phase is preferred since its entropy is smaller than that of the trivial phase, as shown in Fig.~\ref{fig:RT}. Thirdly, when $\lambda$ grows, the first term in (\ref{EECandidate}) also grows with $\lambda$. At the same time, the RT surface $X_p$ in the island phase will stretch into the bulk in order to alleviate the growth of the first term in (\ref{EECandidate}). Fourthly, when the scale of the RT surface $X_p$ is large enough at some values of $\lambda$, the finite width $x_b$ will become negligible. So we can consider the limit $x_b/z_*\to0$ in (\ref{RTAdS}) and find the ratio $\gamma=z_0/z_*\in [0,1]$, which only vanishes at $\theta=0,\pi/2$, as shown in Fig.~\ref{fig:gammalambda}. We can further calculate $\zeta(\rho_0)/\zeta'(\rho_0)$ and $\lambda$ from (\ref{CoordinateRelation}) and (\ref{BC}) at this limit. The value of $\lambda$ at this limit, denoted as $\lambda_c$, is the upper bound on the $\lambda$ for the island phase. We plot $\lambda_c$ as a function of $\theta$ in Fig.~\ref{fig:gammalambda} and find $n|\lambda_c|\leq1$ always. At this limit, $s_p\to1/(n\epsilon^{n})$ approaches the value in the trivial phase from below. Fifthly, when $\lambda\geq\lambda_c$, the island phase does not exist and the RT surface $X_p$ becomes the trivial phase.

\subsection{EWCS ending on the brane}
After figuring out the RT surface of the region $p$, now it is straightforward to define the QEWCS between the subregion $a$ and $b$. Thanks to the translational invariance along $y$ directions, the EWCS between $a$ and $b$ is described by
\begin{align}
E_{a:b}=\ke{(\rho,\zeta,y)|-\infty<\rho<\rho_0,\ 0<\zeta<\zeta_c(\rho),\ y=0},
\end{align}
which intersects with the brane $Q$ at
\begin{align}
    E_{A:B}=\ke{(\rho,\zeta,y)|\rho=\rho_0,\ 0<\zeta<\zeta_c(\rho),\ y=0}.
\end{align}
According to the proposal, the reflected entropy $S^R(a:b)$ can be evaluated by the minimal area of the QEWCS. Therefore, we have 
\begin{align}
S^R(a&:b)=\frac{\lambda L\Area(E_{A:B})+\Area(E_{a:b})}{4G^{(d+1)}}.
\end{align} 

\begin{figure}
    \centering
    \includegraphics[height=0.35\linewidth]{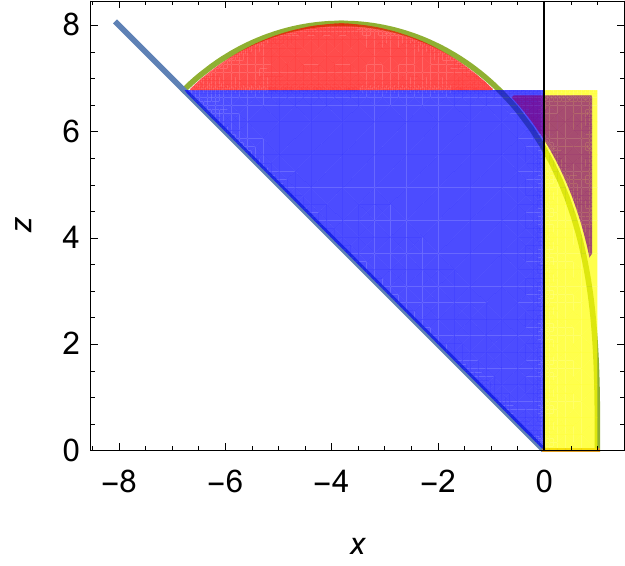}
    \caption{The geometric interpretation of each term in the final expression of the reflected entropy (\ref{REAdS}). The first term corresponds to the yellow rectangle. The second term corresponds to the blue triangle and its bottom margin. The third term corresponds to the red region and the subtraction with the purple region.}
    \label{fig:RERegion}
\end{figure}

When $\lambda<\lambda_c$, {\it i.e.} the island pahse, the density of the reflected entropy is
\begin{align}
s^R_{a:b}
=&\frac{4G^{(d+1)}}{L^{n+1} V_{n-1}}S^R(a:b) \nn \\
=&\lambda \int_{\zeta_\epsilon}^{\zeta_0} d\zeta \kc{\frac{\cosh\rho_0}{\zeta}}^{n}
+\int_\epsilon^{z_0} dz \frac{x_+(z)+z\sinh\rho_0}{z^{n+1}}
+\Theta(\sigma)\int_{z_0}^{z_*} dz \frac{x_+(z)-x_-(z)}{z^{n+1}} \label{REAdS} \\
=&
\begin{cases}
x_b(\epsilon^{-1}-z_0^{-1})+(\lambda  \cosh\rho_0+\sinh \rho_0) \ln (z_0/\epsilon)
+f_1(z_0/z_*,\Theta(\sigma)),& n=1\\
\frac{1}{n}x_b(\epsilon^{-n}-z_0^{-n})+\frac{1}{n-1}(\lambda  \cosh\rho_0+\sinh \rho_0)\kc{\epsilon^{1-n}-z_0^{1-n}}
+z_0^{1-n}f_n(z_0/z_*,\Theta(\sigma)),& n\geq2
\end{cases} \nn
\end{align}
where $\Theta(\sigma)$ is the step function, and $f_n(z_0/z_*,\Theta(\sigma))$ are some complicated functions.  The cutoff is chosen as $z=\epsilon$ and constant $\zeta_\epsilon=\epsilon\cosh\rho_0$. The dependence of the reflected entropy density $s^R_{a:b}$ on $\lambda$ is shown in Fig.~\ref{fig:RT}. We remark that each term in the final expression has its own geometric correspondence and we demonstrate this in Fig.~\ref{fig:RERegion}. Now we elaborate our understanding on these terms as follows. 

The first term roughly measures the entanglement of the bath CFT$_d$ between $a:b$ and thus exhibits area law $L^{d-2}x_bV_{d-3}/\epsilon^n$. 
The second term roughly measures the entropy in the $d$-dimensional gravity-plus-CFT theory between $A:B$ and within wedge $\zeta\in[\zeta_\epsilon,\zeta_0]$ on the brane. It exhibits area law with $\Area(E_{A:B})=L^{d-2}V_{d-3}\cosh\rho_0\int_\epsilon^{z_0} dz z^{2-d}$.

Now we turn to the final target that is the reflected entropy of $A:B$. If we send $x_b\to0$, the bipartition $a:b$ becomes $A:B$ exactly and one would expect to  obtain the reflected entropy $S^R(A:B)$. Unfortunately, due to the scaling symmetry of AdS space, the independent length scale of the RT surface $X_p$ is $x_b$. When $x_b\to0$, all the other length scales of the RT surface, such as $\{z_*,z_0,x_0,\zeta_0\}$, in general become vanishing as well. This issue occurs since for pure AdS, the CFT stays in the ground state with long-range correlation. As a consequence, the CFT on the brane $Q$ is highly entangled with the CFT on $M$. From (\ref{QES}), we notice that the QES $X_P$ tends to contain a small $\Sigma_P$ to resist the high entanglement of CFT, which leads to $z_0\to0$ when $x_b\to0$. To cure this issue, one may increase the proportion of the first term in (\ref{QES}). Here we propose the following two prescriptions: increasing the value of $\lambda$ or adding a black hole. The former prescription will be discussed immediately. The later prescription breaks the scaling symmetry and will be considered in the next section.

After all, if we send $x_b\to\epsilon$, the bipartition $a:b$ effectively approaches $A:B$. Meanwhile, to avoid $z_0\sim \epsilon$ one could further choose a large $\lambda$ approaching $\lambda_c$ from below. According to the analysis in last subsection, the RT surface is subject to $z_0=\gamma z_*\gg x_b$, which stretches into the bulk and keeps away from the conformal boundary $M$ even we send $x_b\to\epsilon$. This tendency has been checked numerically in Fig.~\ref{fig:RT}. 

When $\lambda\geq\lambda_c$, {\it i.e.} in the trivial phase, the cross section $E_{a:b}$ becomes the surface $\{(x,y,z)|\, x<0,\,y=0,\, z>-x\tan\theta\}$ and the density of the reflected entropy is 
\begin{align}\label{RELargelambda}
s_{a:b}^R=
\begin{cases}
(\lambda  \cosh\rho_0+\sinh \rho_0) \ln (z_0/\epsilon)\to\infty\ (z_0\to\infty), & n=1\\
\frac{1}{n-1}(\lambda  \cosh\rho_0+\sinh \rho_0)\epsilon^{1-n}, & n\geq2
\end{cases}
\end{align}
which encounters IR divergence for $n=1$.

Let us discuss the reflected entropy from the boundary perspective. In the island phase, the behavior of the reflected entropy reflects the finite correlation length $\xi$ of the conformal defect on $P$. In the presence of QES at $\zeta_0$, the correlation is suppressed by the entanglement between the conformal defect on $P$ and the bath CFT$_d$ on $M$, whose correlation length along $y$ axis scales as $\xi\sim \zeta_0$ \cite{Sully:2020pza}. In other words, the reflected entropy is dominated by the correlation within the smaller region $\ke{(x,y)|0\leq x<x_b,-\xi<y<\xi}$ in the CFT. When $d=3$, it is similar to the situation of Fig.~\ref{fig:EWCS2D}, where $A\cup B$ is a subregion of $P$ with length scaling as $\xi$. So the reflected entropy scales as $\ln(\xi/\epsilon)\sim \ln(z_0/\epsilon)$ \cite{Dutta:2019gen} and agrees with (\ref{REAdS}). 

In the trivial phase, we have $\zeta_0\to\infty$ and the correlation within $P$ is no longer suppressed by the entanglement between the defect $P$ and the bath CFT on $M$. The reason is that the central charge $c_P$ of the defect on $P$ is comparable to the central charge $c_M$ of the bath CFT on $M$, more precisely $c_P/c_M\sim (1+n\lambda)\csc^n\theta$ \cite{Chen:2020uac}. If we neglect the influence of the bath CFT, the defect on $P$ behaves as a CFT$_{d-1}$, whose reflected entropy between its two half spaces scales as (\ref{RELargelambda}) exactly, where $n=d-2$.

\section{The reflected entropy in the black hole background}

At finite temperature, the $(d+1)$-dimensional bulk geometry is a neutral black hole with a brane, which bends toward the interior of the bulk due to the gravity and touches the horizon. Unlike the case of pure AdS space, now the background is characterized by finite parameters $\ke{\theta,\lambda,T}$, and the RT surface in the bulk would not shrink into zero even for $x_b\to0$. On the other side, with the increase of $\lambda$, the QES will approach the horizon. Following the framework proposed in \cite{Almheiri:2019yqk,Almheiri:2019hni,Almheiri:2019psy,Ling:2020laa}, in this section we will numerically construct the black hole background with a brane, and further evaluate the reflected entropy $S^R_{A:B}$ by the minimal cross-section $E_{A:B}$. Its behavior for different $\lambda$ will be analyzed as well.

\subsection{Background}

To discuss the reflected entropy at finite temperature in the doubly holographic setup, the first thing is to construct a black hole background with a brane. In particular, once the backreaction of the brane is taken into account, one usually needs to solve the equations of motion numerically. Such a static background in higher dimensions has previously been investigated by virtue of Einstein-DeTurck method \cite{Headrick:2009pv,Dias:2015nua,Almheiri:2019hni}. In this paper, we consider the specific case of $d=3$, and for later convenience, we introduce a coordinate system $\kc{t,w,r,y}$ by the following transformation
\begin{align}
\frac{w}{1-w}=x+z\cot\theta,\quad r=\sqrt{1-z}.
\end{align}
In this coordinate system, the metric ansatz for a black hole background is given as
\begin{align}\label{eq:Ansatz}
ds^2=&\frac{L^2}{(1-r^2)^2}\left[-r^2 P(r) F_1 dt^2 +\frac{4 F_2}{P(r)}dr^2+
\frac{F_4}{(1-w)^4}\left(dw+ 2 r (1-w)^2 F_3 dr \right)^2+F_5dy^2\right],
\end{align}
where 
\begin{align}\label{eq:P}
P(r)=&2-r^2+(1-r^2)^2,
\end{align}
and $\ke{F_i|i=1,2,...,5}$ are functions of  $\kc{r,w}$ in the domain $\ke{0<w<1,0<r<1}$.

The configuration of the background is given by the following setup. The brane $Q$ is located at $w=0$. The infinity $I$ far from the brane is located at $w=1$. The boundary $M$ is located at $r=1$ and the horizon $H$ is located at $r=0$.

Instead of solving the Einstein equation directly, we will solve the Einstein-DeTurck equations \cite{Almheiri:2019hni,Headrick:2009pv,Dias:2015nua}
\begin{align}\label{eq:EDEquations}
R_{\mu\nu}+3\, g_{\mu\nu}=\nabla_{(\mu}\xi_{\nu)},\quad 
\xi^{\mu}=\left[\Gamma_{\nu \sigma}^{\mu}(g)-\Gamma_{\nu
	\sigma}^{\mu}(\bar{g})\right] g^{\nu \sigma},
\end{align}
where $\xi^\mu$ is the DeTurck vector and $\bar{g}$ is the reference metric.
The boundary conditions are imposed as follows
\begin{equation}
\begin{array}{llllll}
r=1: & F_1=1,  &  F_2=1,  &  F_3=\cot\theta,  &  F_4=1,  &  F_5=1;  \\
r=0: &  \partial_r F_1=0,  &  \partial_r F_2=0,  &  \partial_r F_3=0,  &  \partial_r F_4=0,  &  \partial_r F_5=0; \\
w=1: &  F_1=1,  &  F_2=1,  &  F_3=\cot\theta,  &  F_4=1,  &  F_5=1; \\
w=0: & n_\mu \xi^{\mu}=0, &  F_3=\cot\theta,  &  \multicolumn{3}{l}{\text{Eq.~}(\ref{eq:BCBrane}).}  
\end{array}
\end{equation}
The reference metric $\bar g$ should be subjected to the same boundary conditions as $g$ on the surfaces $\ke{I,H,M}$, thus we choose it to be an AdS-Schwarzschild black hole with
\begin{align}
F_1=1,\quad F_2=1,\quad F_3=\cot\theta,\quad F_4=1,\quad F_5=1.
\end{align}
With the general metric ansatz (\ref{eq:Ansatz}), the background is obtained numerically via the Newton-Raphson method, where we discretize (\ref{eq:EDEquations}) on $r$ and $w$ directions with Chebyshev Pseudo-spectral method.

\begin{figure}
    \centering
    \includegraphics[width=138pt]{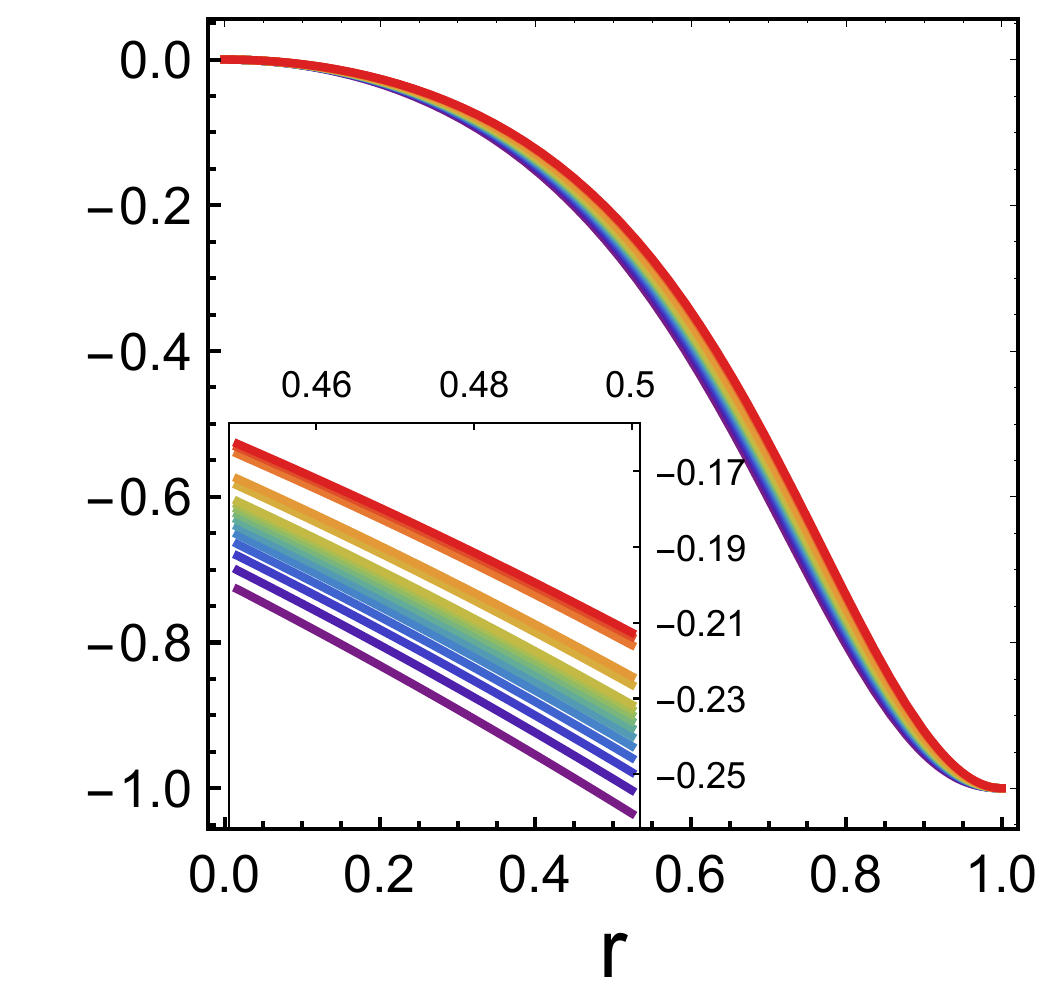}
    \includegraphics[width=127pt]{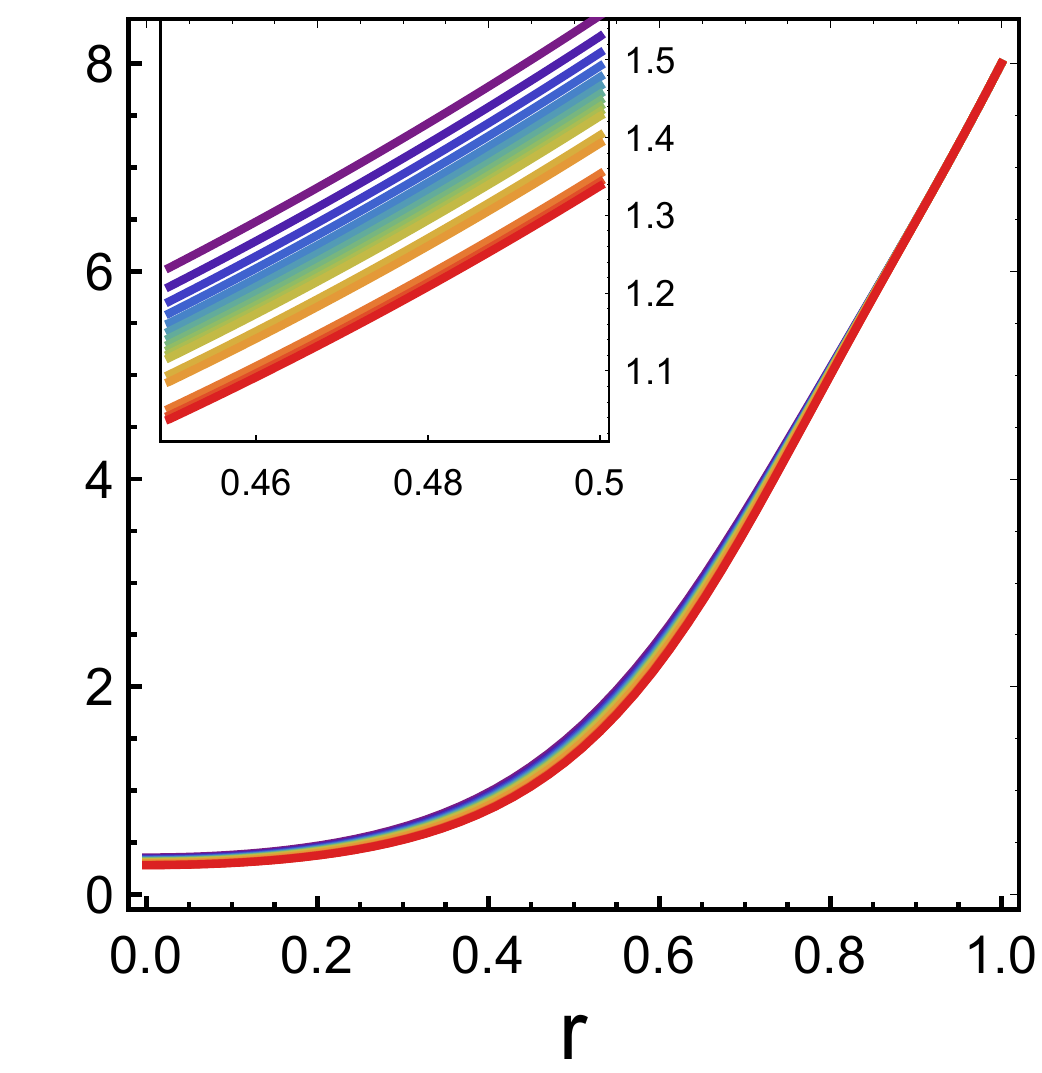}
    \includegraphics[width=133pt]{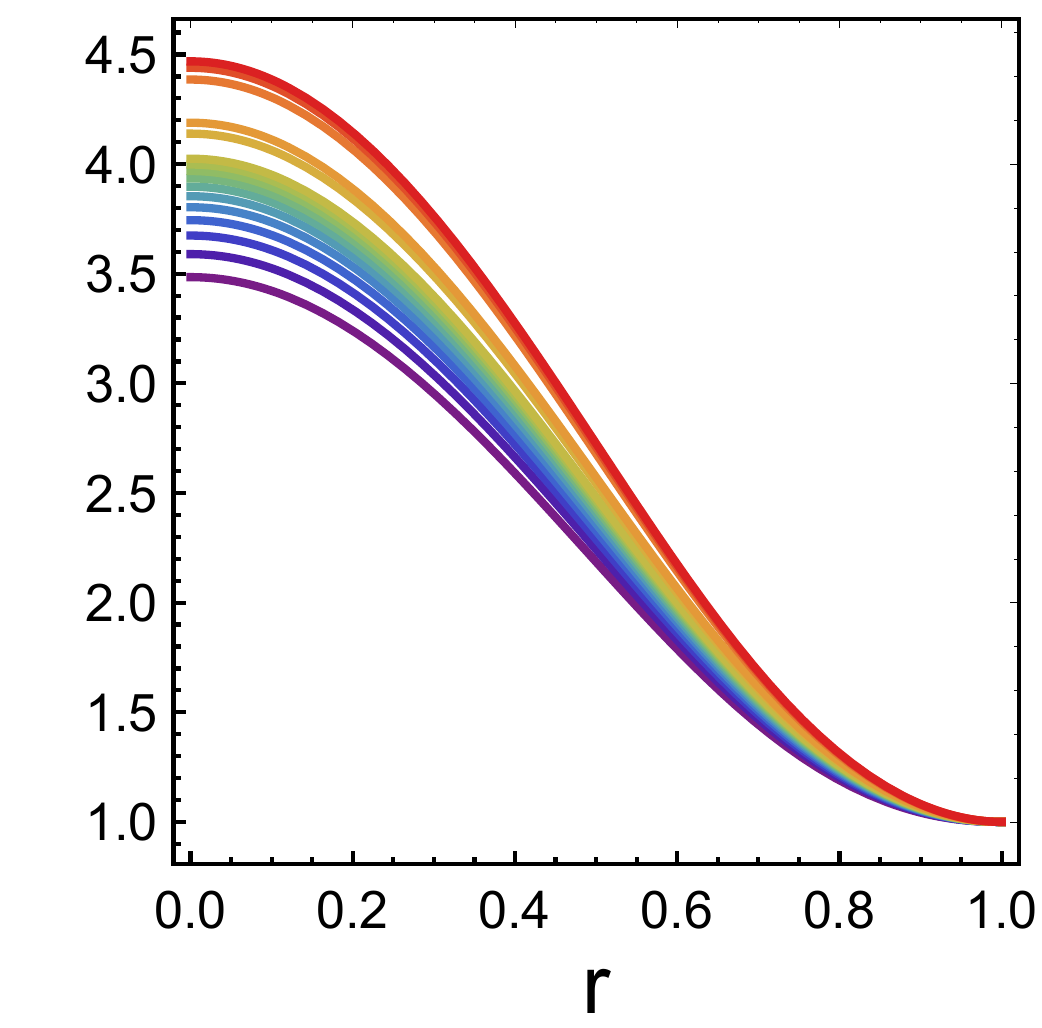}\\
    \includegraphics[width=138pt]{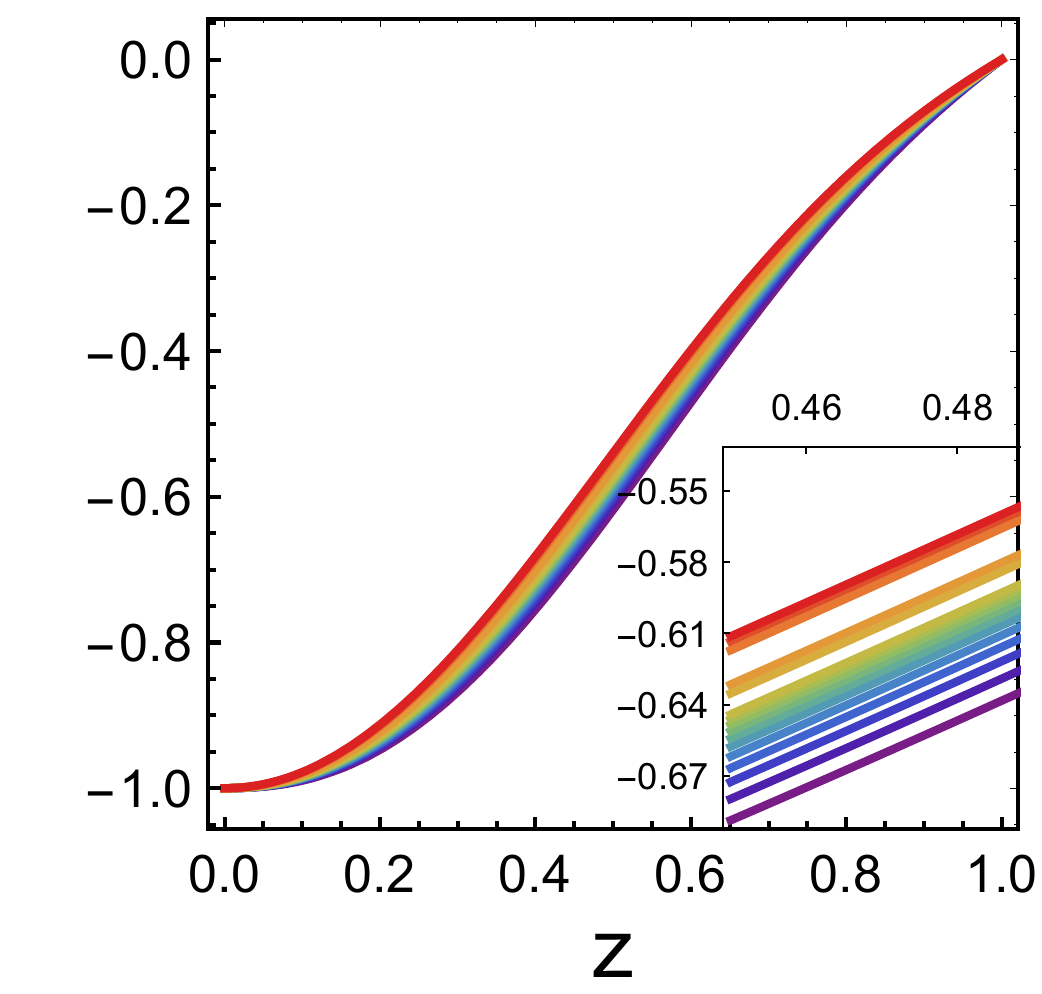}
    \includegraphics[width=133pt]{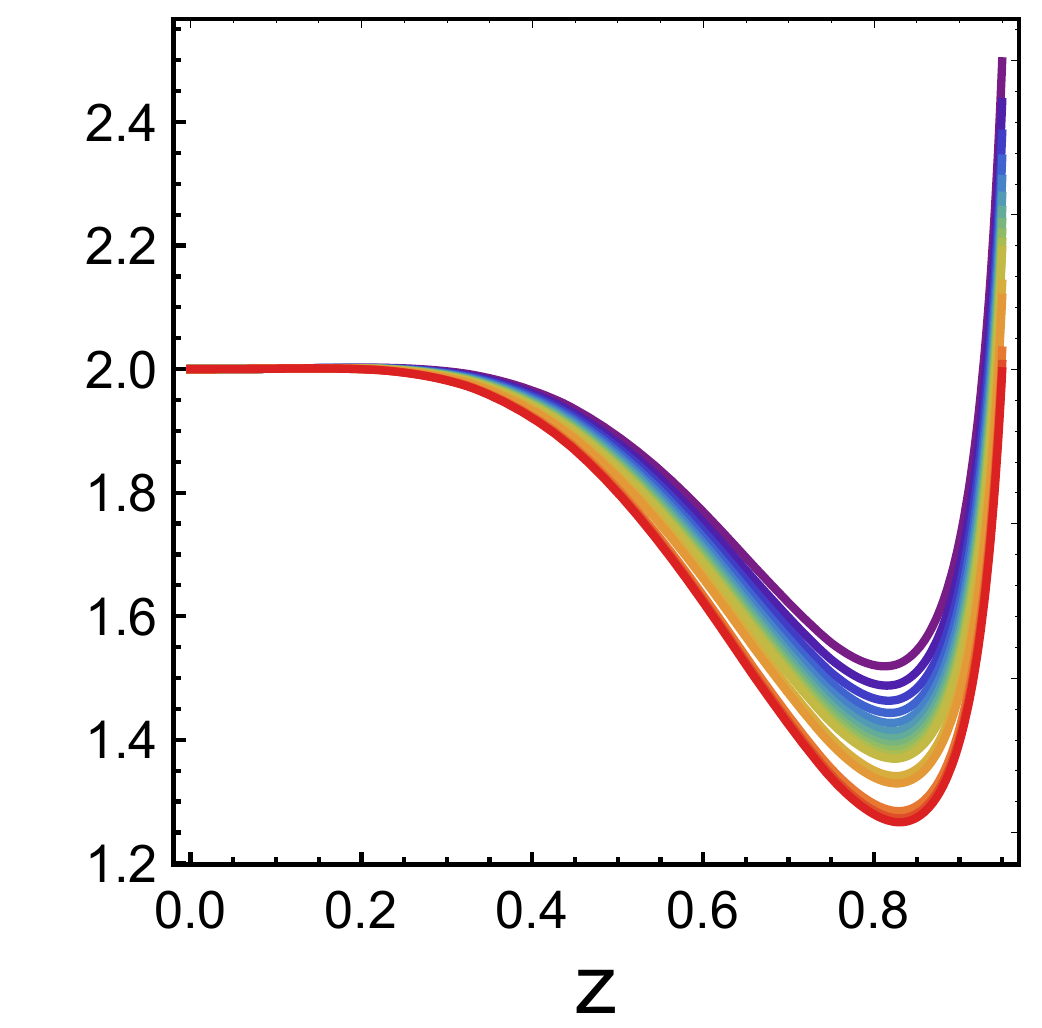}
    \includegraphics[width=133pt]{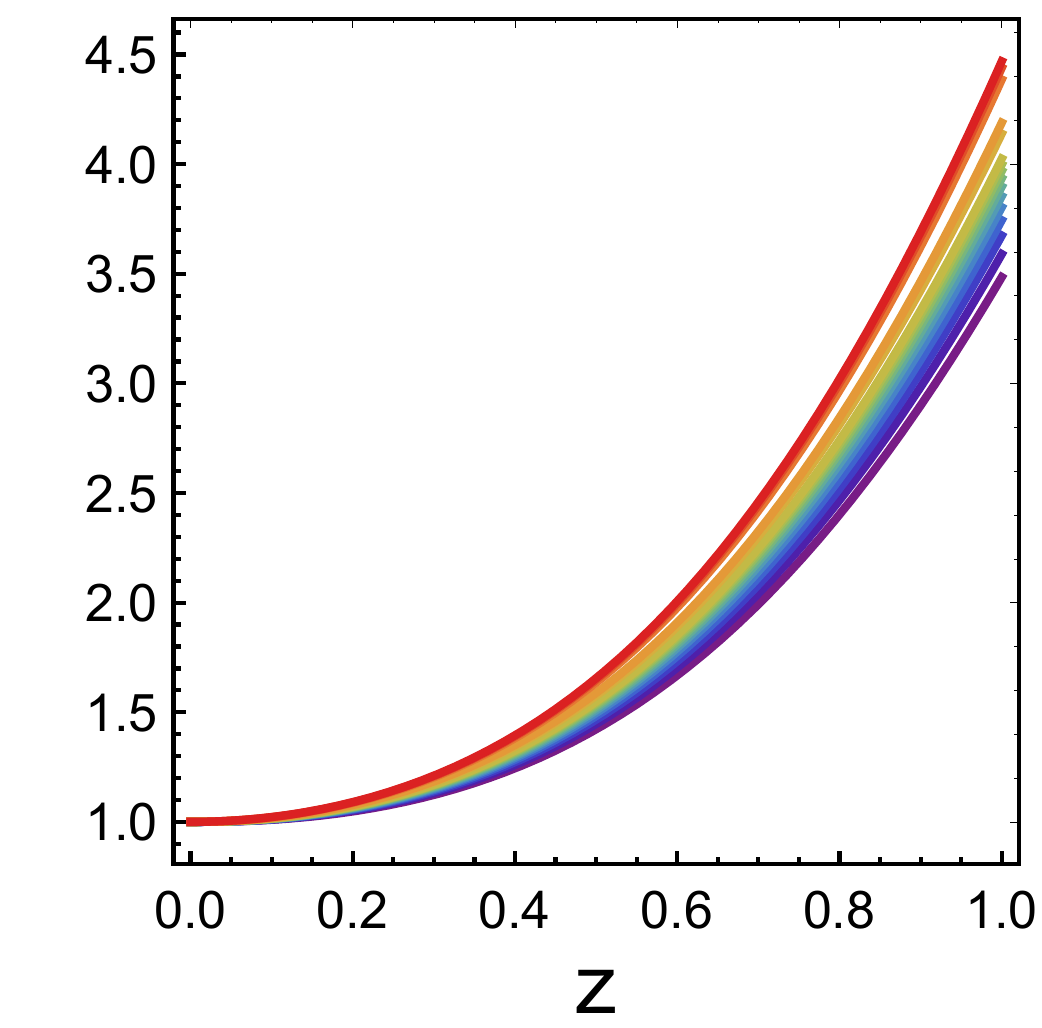}
    \caption{In the first row, three components of the induced metric on the brane, $h_{tt}$, $h_{rr}$ and $h_{yy}$, are depicted respectively in coordinates $(t,w,r,y)$. All the components are multiplied by a factor $\frac{(1-r^2)^2}{L^2}$. While in the second row, three components of the induced metric on the brane, $h_{tt}$, $h_{zz}$ and $h_{yy}$, are depicted in different coordinates $(t,z,x,y)$. All the components are multiplied by a factor $\frac{z^2}{L^2}$. Here $L$ is fixed to be $1$ and the curves from violet to red are plotted with different $\lambda=0.3,0.4,0.5,0.6,0.7,0.8,0.9,1,1.6,2,7,14,28$.}
    \label{fig:MetricComponentsOnBrane}
\end{figure}

To take the backreaction of the brane into account, hereafter, we will fix $\theta=\pi/4$ and vary $\lambda$. With different $\lambda$, the numerical solutions of the induced metric on the brane ($w=0$) are illustrated in Fig.~\ref{fig:MetricComponentsOnBrane}. Note that in the original coordinates $(t,x,y,z)$, the component $h_{zz}$ is divergent on the horizon $z=1$, but the apparent divergence at $r=0$ vanishes in the new coordinates $(t,r,w,y)$.

\subsection{RT surface}

Next, for a given bipartite system, we intend to determine the RT surface over the black hole background. Following the scheme in \cite{Almheiri:2019psy, Ling:2020laa}, we divide the RT surface into two segments by the turning point, and each of which can be parameterized by
\begin{align}
\kc{r,w}=
\begin{cases}
\kc{r(w),w},& 0\leq w<w_t\\
\kc{r,w(r)},& r_t\leq r\leq 1
\end{cases}
\end{align}
respectively. Then the entropy density is the minimum of the sum of two area terms
\begin{align}
\tilde s_p
= \frac{\lambda L \Area(\tilde X_P)+\Area(\tilde X_p)}{L^2V_1},
\end{align}
where
\begin{align}
(\tilde s_p)_{bulk}:= & \frac{\Area(\tilde X_p)}{L^2V_1}\\ 
= &\int_{r_t}^{r_\epsilon} \frac{dr}{\left(1-r^2\right)^2} \sqrt{F_5 \left(\frac{4F_2}{P(r)}+\frac{F_4 \left(2 r (w(r)-1)^2 F_3+w'(r)\right)^2}{(w(r)-1)^4}\right)} +\int_{0}^{w_t} \frac{dw}{\left(r(w)^2-1\right)^2}
\nn\\
&\sqrt{F_5 \left(\frac{4 F_2 r'(w)^2}{P(r(w))}+F_4 \left(4 F_3^2 r(w)^2 r'(w)^2+\frac{4 F_3 r(w) r'(w)}{(w-1)^2}+\frac{1}{(w-1)^4}\right)\right)},\label{eq:SBulk}\\
(\tilde s_p)_{DGP}:= & \frac{\lambda L \Area(\tilde X_P)}{L^2V_1} 
=\frac{\lambda \sqrt{F_5}}{1-r(0)^2}\Big|_{w=0}, \label{eq:SBrane}
\end{align}
with $r_\epsilon=\sqrt{1-\epsilon}$.

$\tilde s_p$ as well as the intersection $r= r(0)$ changes with the shift of the turning point $\ke{r_t,w_t}$. When $\tilde s_p$ reaches its minimum $ s_p$, we denote the corresponding solutions as  $r^c (w)$ and $w^c (r)$ for each segment, and the intersection as $r=r^c(0)$.

\begin{figure}
    \centering
    \includegraphics[height=0.3\linewidth]{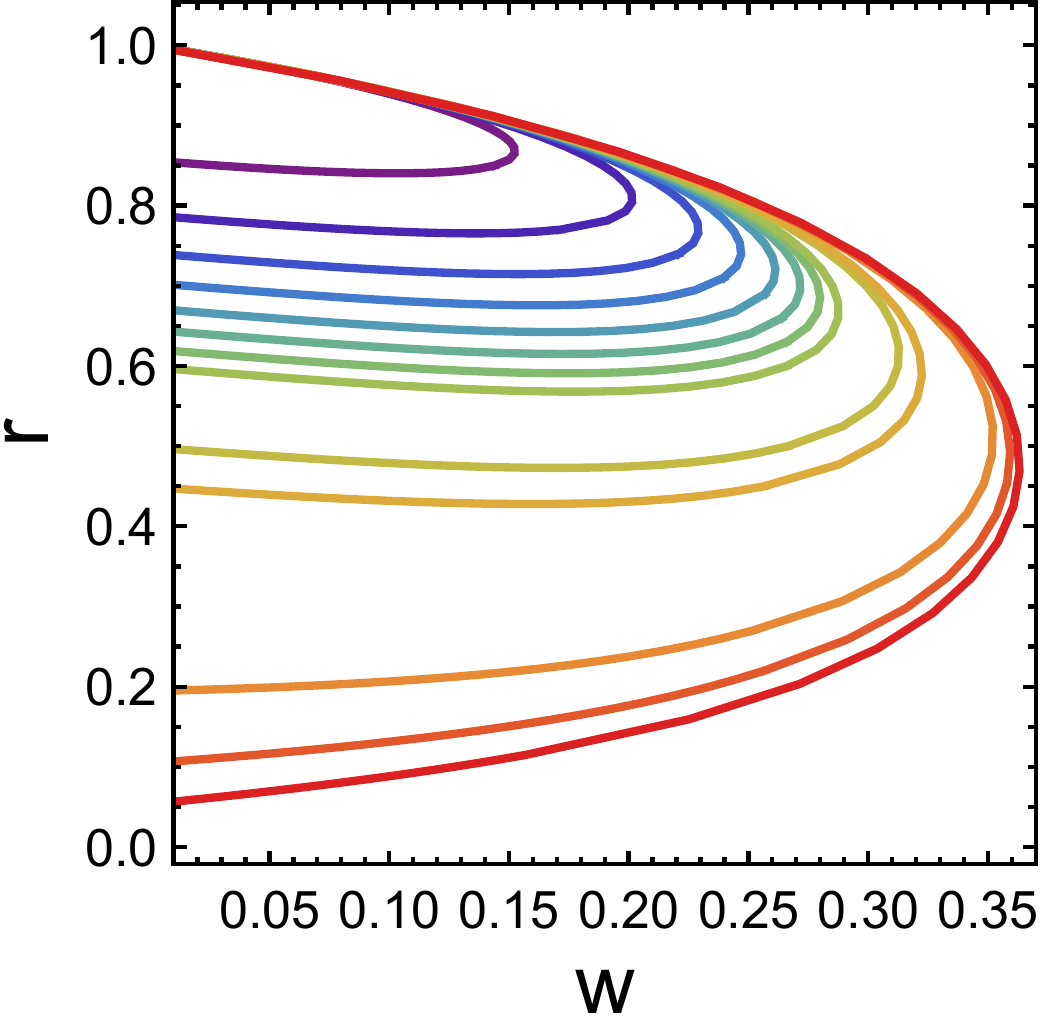}
    \includegraphics[height=0.3\linewidth]{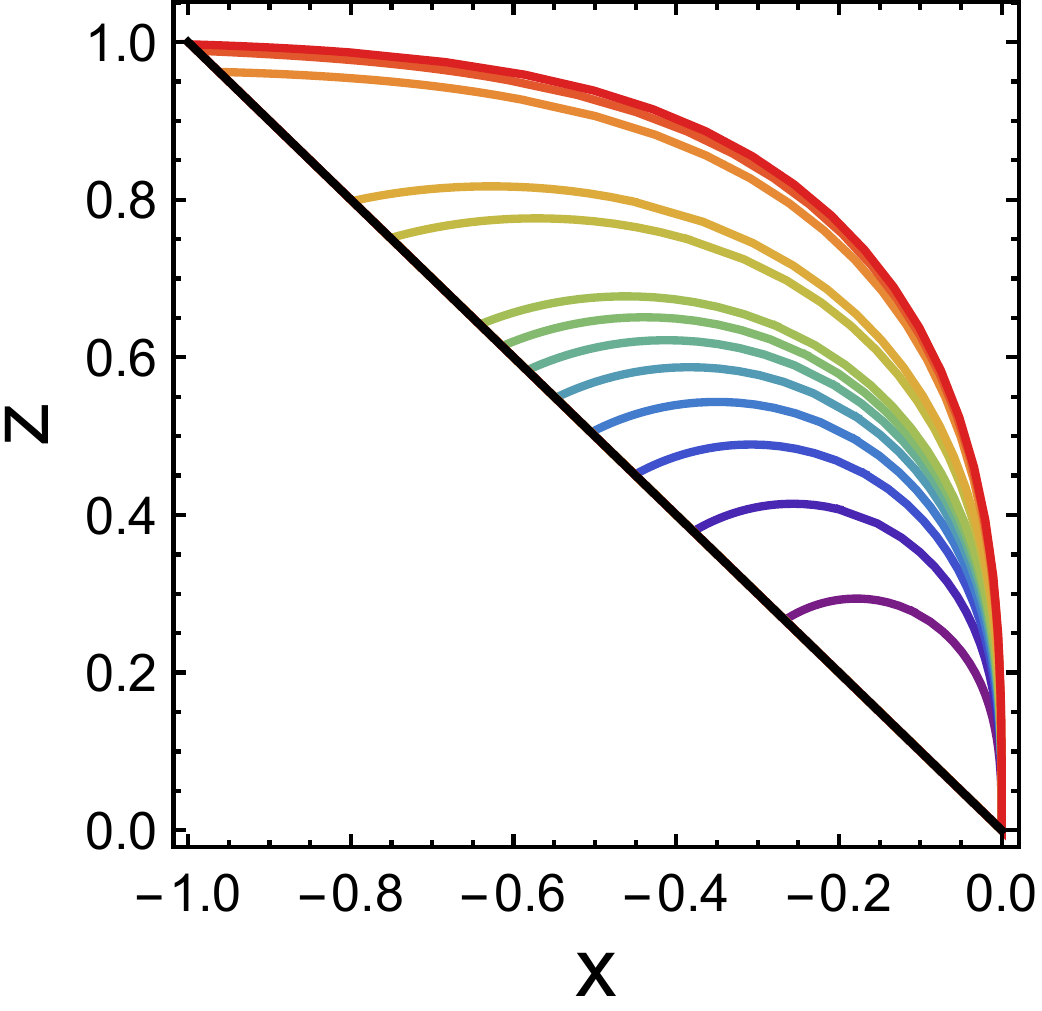}
    \caption{The RT surfaces in coordinates $(t,r,w,y)$ and $(t,z,x,y)$ for different $\lambda$ in $4$-dimensional black hole, where $\{d,\theta,x_b\}=\{3,\pi/4,0\}$ and $\lambda=0.3,0.4,0.5,0.6,0.7,0.8,0.9,1,1.6,2,7,14,28$ (from the UV region in violet to the IR region in red).}
    \label{fig:RTSurfaces}
\end{figure}

\begin{figure}
    \centering
    \includegraphics[height=0.35\linewidth]{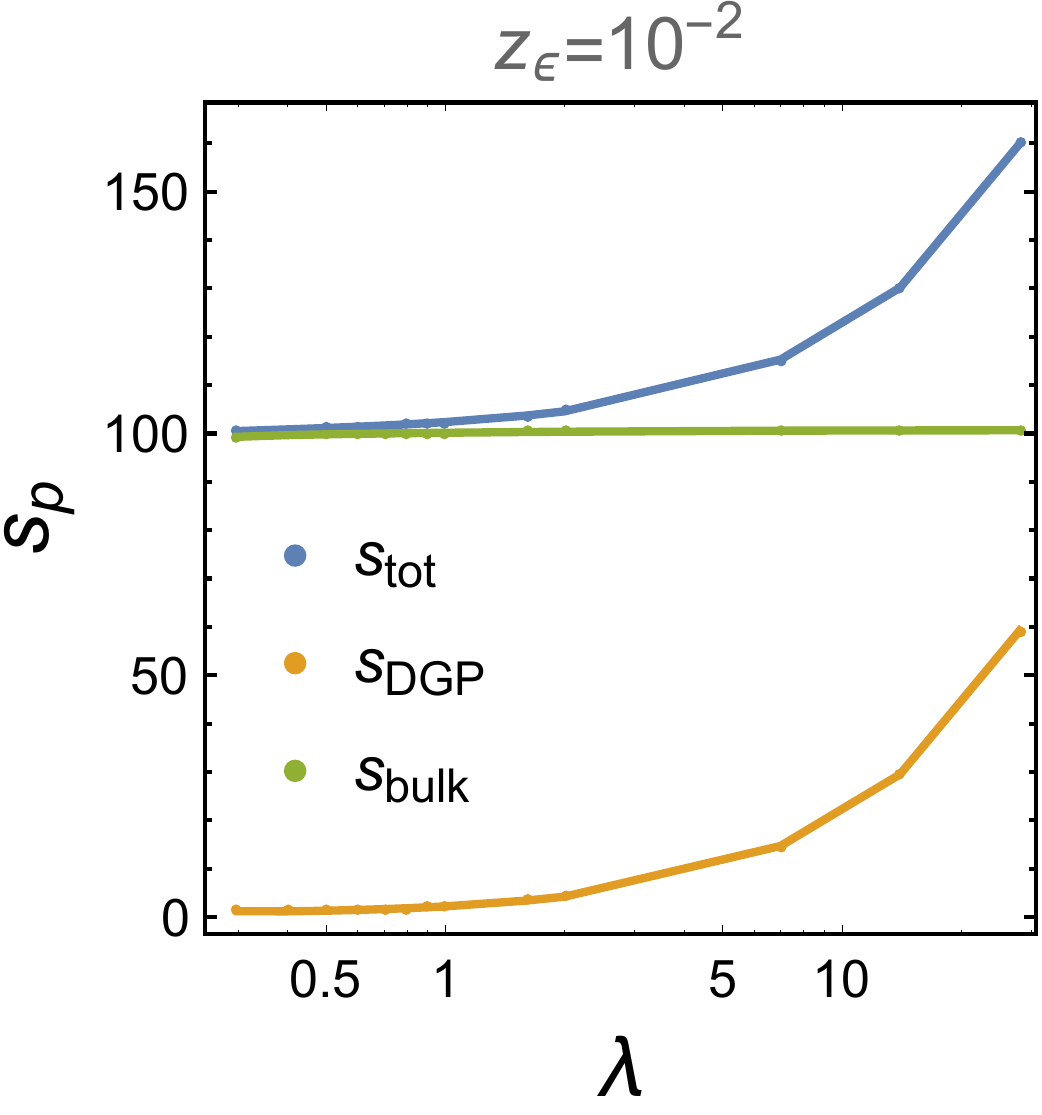}
    \includegraphics[height=0.35\linewidth]{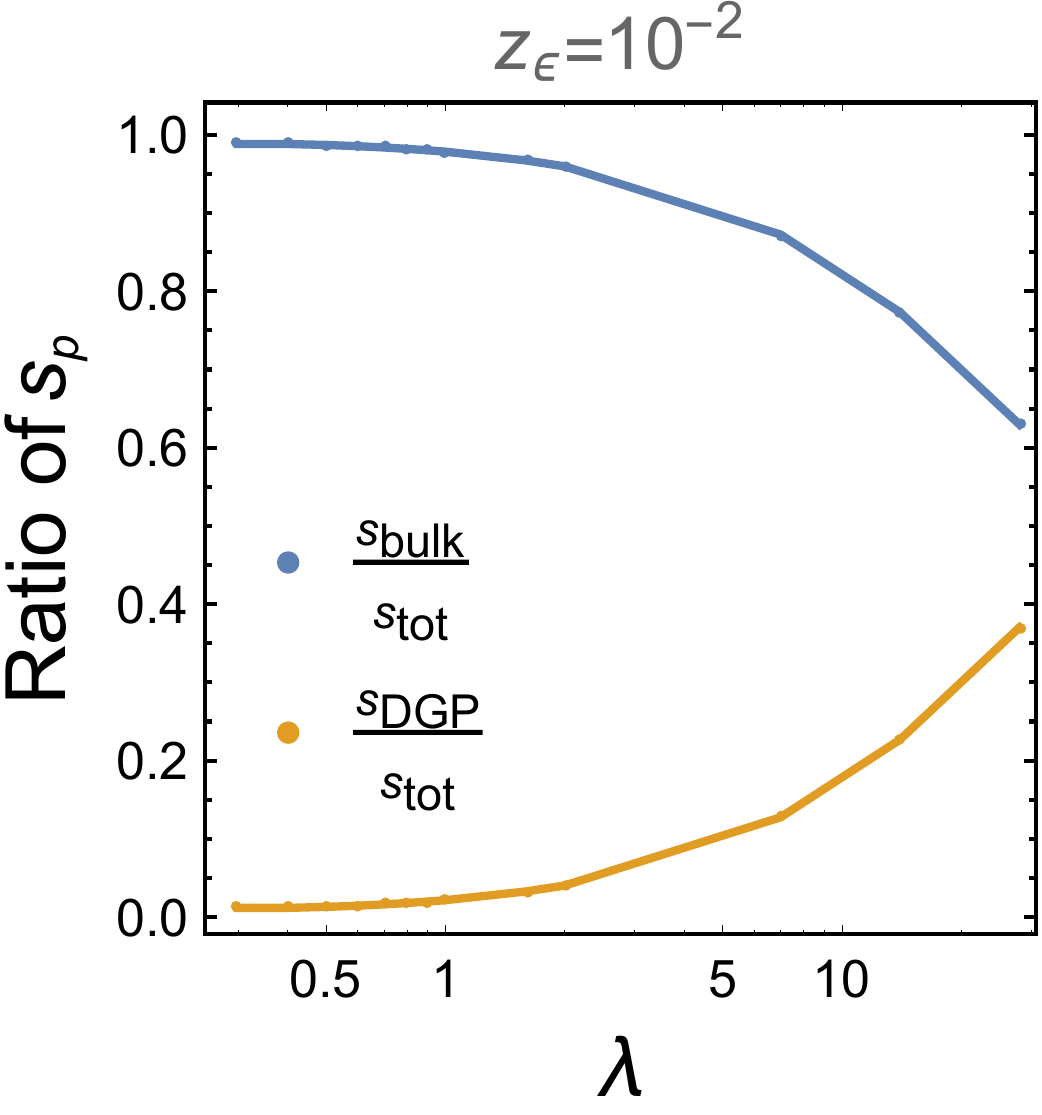}
    \caption{The entanglement entropy density $s_p$ and the contributions of each component $(s_p)_{DGP}$ and $(s_p)_{bulk}$ for different $\lambda$, where the UV cut-off is $\epsilon=0.01$.}
    \label{fig:spOfLambda}
\end{figure}

In the limit $x_b\to0$, the region $p$ becomes $P$, and the corresponding entropy $S_p$ becomes $S_P$, which is the quantity we are really concerned with. The configurations of RT surfaces for different values of $\lambda$ are shown in Fig.~\ref{fig:RTSurfaces}. Since the black hole breaks the scaling symmetry of AdS$_4$ space, in general the RT surface no longer shrinks to the boundary of the brane. For small $\lambda$, the RT surface is located near the boundary, and the configuration is similar to the vacuum case; while for large $\lambda$, the configuration of the RT surface is stretched along $x$ direction near the horizon. Moreover, with the growth of $\lambda$, the increase of the entanglement entropy density $s_P$ is almost contributed from the increase of $(s_P)_{DGP}$, while $(s_P)_{bulk}$ is almost a constant, as shown in Fig.~\ref{fig:spOfLambda}.

This tendency can also be understood from the brane perspective. In the presence of a black hole at finite temperature, the CFT$_3$ on $Q\cup M$ is characterized by the correlation with finite length. When one searches for the QES of $P$ by utilizing (\ref{QES}), the entropy density of the CFT within the wedge $\Sigma_P$ is bounded by the thermal correlation length. For a large $\lambda$, {\it i.e.} a small $G^{(d)}$, the entropy is dominated by the geometric term in (\ref{QES}). Therefore,  the QES approaches the horizon.

\subsection{EWCS ending on the brane}

Now we consider the reflected entropy between two subsystems by evaluating the area of the EWCS. As $x_b\to0$, the bipartition $a:b$ becomes $A:B$ exactly and the reflected entropy $S^R(a:b)$ becomes $S^R(A:B)$. The reflected entropy is contributed by the area of the $2$-dimensional cross-section $E_{a:b}$ in the bulk $N$ and the area of the $1$-dimensional cross-section $E_{A:B}$ on the brane $Q$, namely,
\begin{align}
S^R(A:B)=&\frac{\lambda L\Area(E_{A:B})+\Area(E_{a:b})}{L^2}, 
\end{align}
where
\begin{align}
[S^R(A:B)]_{bulk}:=&\frac{\Area(E_{a:b})}{L^2}\\
=&\kc{\int_{r_t}^{r_\epsilon} dr \int_{\frac{1-r}{1-r_t}w_t}^{w^c(r)} dw + \int_0^{w_t} dw \int_{r^c(w)}^{\min\kd{r_\epsilon,1-\frac{w}{w_t}(1-r_t)}} dr} \frac2{(1-r^2)^2(1-w)^2}\sqrt{\frac{F_2F_4}{P(r)}}	\nn \\
[S^R(A:B)]_{DGP}:=&\frac{\lambda L\Area(E_{A:B})}{L^2}
=\lambda \int_{r^c(0)}^{r_\epsilon} dr  \frac{2}{1-r^2}\sqrt{\frac{F_2}{P(r)}+r^2F_3^2F_4}.
\end{align}
Similarly, the cross-section $E_{a:b}$ is parameterized into two parts by the line $w/w_t=(1-r)/(1-r_t)$.

\begin{figure}
	\includegraphics[height=0.35\linewidth]{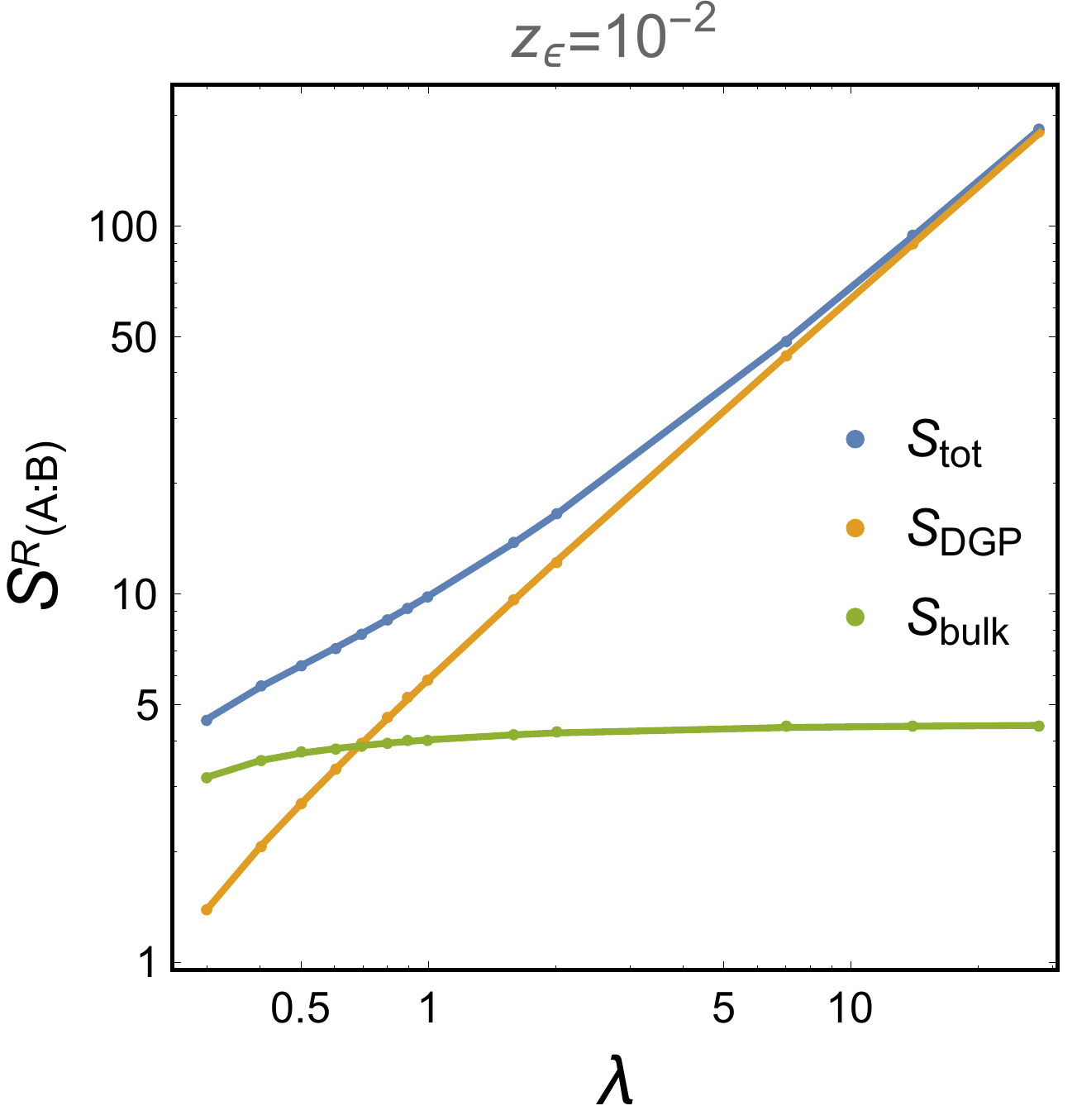}
	\includegraphics[height=0.35\linewidth]{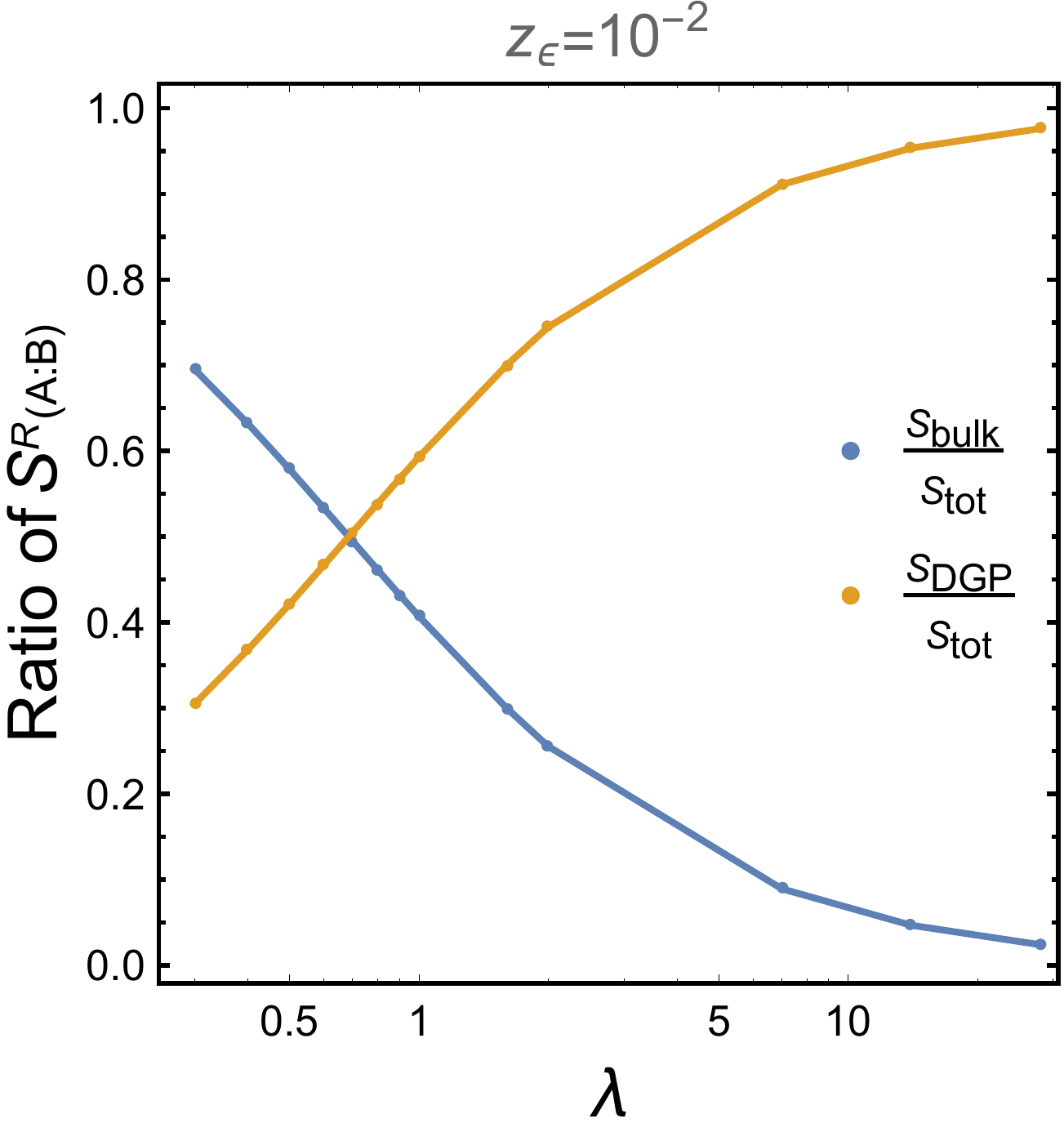}
	\caption{The reflected entropy $S^R(A:B)$ and the contributions from each component $[S^R(A:B)]_{DGP}$ and $[S^R(A:B)]_{bulk}$ for different $\lambda$, where the UV cut-off is $r_\epsilon=\sqrt{1-\epsilon}=\sqrt{1-10^{-2}}$.}
	\label{fig:SofL}
\end{figure}

The numerical results for the reflected entropy and its behavior with the change of $\lambda$ are illustrated in  Fig.~\ref{fig:SofL}. Similar to the case of entanglement entropy, with the growth of $\lambda$, the increase of the reflected entropy $S^R(A:B)$ is mainly contributed from the increase of $[S^R(A:B)]_{DGP}$, while the $[S^R(A:B)]_{bulk}$ grows tardily.

From the brane perspective, due to the finite length of the thermal correlation on the brane, the reflected entropy contributed by the CFT in (\ref{QEWCS}) is bounded above. For large $\lambda$, the geometric term becomes dominant. With the increase of $\lambda$, the geometry on the brane changes slowly, as shown in Fig.~\ref{fig:MetricComponentsOnBrane}, and so does the area of $E_{A:B}$. While $1/G^{(d)}$ increases linearly for large $\lambda$. So the reflected entropy $S^R(A:B)$ increases linearly, as shown in Fig.~\ref{fig:SofL}.

\section{Conclusion and outlook}

In this paper, we have investigated the reflected entropy including the entanglement of quantum matter via the doubly holographic setup. We have proposed a notion of quantum entanglement wedge cross-section (QEWCS), which minimizes the sum of the geometric contribution and quantum matter contribution in (\ref{QEWCS}), and may describe the reflected entropy with higher-order quantum corrections. Specifically, we have considered a $(d+1)$-dimensional gravity theory in AdS with a brane anchoring on the conformal boundary, which is dual to the gravity-plus-CFT theory living on the brane and the bath CFT living on the conformal boundary. Taking the tension and DGP term on the brane into account, we have obtained the reflected entropy between a bipartition of the boundary in the gravity-plus-CFT theory by calculating the minimal area of the corresponding entanglement wedge cross-section in the $(d+1)$-dimensional space. In general, the reflected entropy consists of two parts, one contributed by the geometry on the brane and the other contributed by the CFT on the brane. We have computed their proportion for different Newton constants in the DGP term and found that their behavior agrees with the analysis based on semi-classical gravity and the correlation of CFT coupled to the bath CFT.

It is worthwhile to point out that due to the parity $y\to-y$ of the bipartition $A:B$ chosen in this paper, the configurations of the QEWCS in (\ref{QEWCS}) and the EWCS in (\ref{EWCS}) happen to be the same. Nevertheless, we intend to stress that their definitions are quite different. It is worth further studying the QEWCS of the bipartition without parity in further, and it is expected that the configurations of QEWCS and EWCS should be different.

The reflected entropy in double holography gives a way to compute the entanglement contributed from quantum matter in the bulk of spacetime. Our setup may also be applied to an eternal black hole coupled to the baths, which recently plays a key role in the understanding of the black hole information loss paradox \cite{Almheiri:2019yqk,Almheiri:2019hni,Dias:2015nua,Chen:2020uac,Chen:2020hmv,Geng:2020qvw}.

\section*{Acknowledgments}
We are grateful to Cheng Peng, Shao-Kai Jian for helpful discussions. 
This work is supported in part by the National Natural
Science Foundation of China under Grant No.~11875053, 12075298, 12035016, 11805083, 11905083, 12005077, 11947067 and Guangdong Basic and Applied Basic Research Foundation under Grant No.~2021A1515012374.
Zhuo-Yu Xian also acknowledge support from the Deutsche Forschungsgemeinschaft (DFG, German Research Foundation) under Germany’s Excellence Strategy through the W\"urzburg-Dresden Cluster of Excellence on Complexity and Topology in Quantum Matter ct.qmat (EXC 2147, project id 390858490).
Liu Yuxuan acknowledges the support from the National
Postdoctoral Program for Innovative Talents BX2021303, funded by
China Postdoctoral Science Foundation.

\bibliography{refs}
	
\end{document}